\documentclass[fleqn,usenatbib]{mnras}

\usepackage{amsmath}

\usepackage{newtxtext,newtxmath}

\usepackage[T1]{fontenc}

\DeclareRobustCommand{\VAN}[3]{#2}
\let\VANthebibliography\thebibliography
\def\thebibliography{\DeclareRobustCommand{\VAN}[3]{##3}\VANthebibliography}

\usepackage{graphicx}	
\usepackage{amsmath}	
\newcommand\textlcsc[1]{\textsc{\MakeLowercase{#1}}}

\title[Did GES form the Milky Way's bar?]{Did the Gaia Enceladus/Sausage merger form the Milky Way's bar?}

\author[A. Merrow et al.]{
Alex Merrow,$^{1}$\thanks{E-mail: a.j.cooke@2022.ljmu.ac.uk}
Robert J. J. Grand,$^{1}$
Francesca Fragkoudi,$^{2,3}$
Marie Martig,$^{1}$
\\
$^{1}$Astrophysics Research Institute, Liverpool John Moores University, 146 Brownlow Hill, Liverpool, L3 5RF, UK\\
$^{2}$Department of Physics, Durham University, South Road, Durham DH1 3LE, UK\\
$^{3}$Institute of Computational Cosmology, Department of Physics, Durham University, South Road, Durham DH1 3LE, UK
}

\date{Accepted XXX. Received YYY; in original form ZZZ}

\pubyear{2023}

\begin{document}
\label{firstpage}
\pagerange{\pageref{firstpage}--\pageref{lastpage}}
\maketitle

\begin{abstract}
The Milky Way's last significant merger, the Gaia Enceladus/Sausage (GES), is thought to have taken place between $8-11\,\mathrm{Gyr}$ ago. Recent studies in the literature suggest that the bar of the Milky Way is rather old, indicating that it formed at a similar epoch to the GES merger. We investigate the possible link between these events using one of the Auriga cosmological simulations which has salient features in common with the Milky Way, including a last significant merger with kinematic signatures resembling that of the GES. In this simulation, the GES-like merger event triggers tidal forces on the disc, gas inflows and a burst of star formation, with the formation of a bar occuring within $1\,\mathrm{Gyr}$ of the first pericentre. To highlight the effects of the merger, we rerun the simulation from $z=4$ with the progenitors of the GES-like galaxy removed well before the merger time. The consequence is a delay in bar formation by around $2\,\mathrm{Gyr}$, and this new bar forms without any significant external perturbers. We conclude that this Milky Way-like simulation shows a route to the real Milky Way's bar forming around the epoch of the GES merger due to tidal forces on its first pericentre. We explore all Auriga galaxies with GES-like merger events, and find that those with stellar mass ratios below $10\%$ form bars within $1\,\mathrm{Gyr}$ of the merger, while bar formation is delayed in the more massive merger scenarios. These include the 4 oldest bars in the simulation suite. Lastly, we note some later morphological differences between the disc of the original simulation and our rerun, in particular that the latter does not grow radially for the final $7\,\mathrm{Gyr}$. Our study suggests that the GES may therefore be responsible for the formation of the Milky Way's bar, as well as for the build-up of its extended disc.
\end{abstract}

\begin{keywords}
methods: numerical -- galaxies: bar -- Galaxy: disc -- Galaxy: evolution -- galaxies: interactions -- Galaxy: kinematics and dynamics
\end{keywords}



\section{Introduction}
One of the most striking features observed in present-day disc galaxies is bars in their central regions, structures which are prevalent across the local Universe. \citet{2018Er} observes bars in around 70\% of nearby disc galaxies \citep[in agreement with e.g.][]{2007En,2008Sh} and finds that this bar fraction is primarily dependent on stellar mass. \citet{2011Ma} and \citet{2014Me} find a lower proportion of bars, around 30\%, in tension with this fraction, but \citet{2011Ma} notes that this difference could easily be accounted for by the differing methods of bar detection. We observe fewer bars in younger galaxies at higher redshifts \citep{2008Sh}; however, the James Webb Space Telescope has uncovered bars up to redshift $\sim3$ \citep{2022Gu,2023Co,2023Cs}, suggesting that this anti-correlation of bar fraction with redshift may be challenged as observations continue to improve. Across a wide variety of disc galaxy populations, however, we observe in each a significant number without any bar-like structures, with no comprehensive correlation from any of the galaxies' properties. This suggests that some mechanisms of bar formation occur in specific cases only, but are widespread across the Universe and cosmic time.

Simulations offer a view of a galaxy across these huge time scales, giving us a view of its bar's full evolution and in particular its formation. Indeed, simulations of isolated disc galaxies have long been known to form bars \citep[e.g.][]{1971Ho,1972Ka,2002At2}. The prevalent formation of bars in a wide variety of idealised simulations indicates that bars certainly can form in isolation due to gravitational instabilities in the disc, trapping stars in bar-like orbits, even if some of the exact mechanisms are still debated.  Further simulations introduce interactions to the model, for example \citet{1987No,1990Ge,1998Mi,2014Lo,2017Ma} all show the formation of bars by tidal interactions from a second perturbing galaxy. With these simulations, we have two methods of bar formation: internal instabilities and tidal interactions. Mergers both produce tidal forces on their approach and introduce instabilities to the disc either upon impact or by adding to the galaxy's mass, creating complex scenarios where the exact cause of bar formation remains unclear. \citet{2016At} shows that bars can form immediately after major mergers of galaxies, whereas \citet{2021Gh} shows that bars can instead be weakened by mergers in many cases, or even destroyed as in \citet{2013Gu} \citep[although simulations have shown that bars generally are long-lived structures e.g.][]{1987Sp,2003At,2012Kr}.

Zoom-in cosmological simulations sacrifice precise control over a galaxy's initial conditions in return for a realistic merger history, accretion of gas, and evolution within the cosmological context. In addition, by reducing the resolution of the wider simulation box, these simulations can benefit too from the computing space for high dynamical resolution in addition to the more complex interactions between their matter, including magnetic fields and models for stellar feedback. This combination of details leads to the possibility of disentangling the processes involved in a merger, while still preserving a wide variety of interactions. Simulations from \citet{2022Bi} for example contain an overwhelming prevalence of barred galaxies before redshift 2, all of which have their formation triggered by external interactions. \citet{2022Ro} also contains more bars than typically observed at higher redshifts, contradicting observations. In contrast, \citet{2012Kr} supports a low bar fraction before redshift 1, citing mergers as the main factor disrupting bars. Uncovering the true effects of various types of merger on bar formation appears key to resolving this tension.

The Milky Way itself has a strong bar \citep{1975Pe,1992We,1999We}, and despite our obscuring viewing angle from within the galaxy we can determine many of the bar's properties, including a pattern speed of $\sim40\,\mathrm{km}\,\mathrm{s}^{-1}\,\mathrm{kpc}^{-1}$ \citep{1999We,2008Ro,2015SM,2019Sa} and a length of $3-5\,\mathrm{kpc}$ dependent on the inclusion of possible spiral-linked structures at the bar's ends \citep{2002Bi,2006Lo,2012Ro,2015We}.

The bar's age is highly debated, due to bars' present properties not having a strong dependence on formation time. \citet{2019Bo} examine the formation history of the Milky Way's stars across the bar and disc, and find that the bar's population is significantly older than the disc's, from which they conclude that the bar formed around $8\,\mathrm{Gyr}$ ago. Oxygen-rich stars in particular trace the shape of the bar for populations as old as $9\,\mathrm{Gyr}$, leading \citet{2020Gr} to conclude that this is when the bar formed. In contrast, \citet{2002Co} find a population of intermediate age, carbon-rich stars tracing the bar and suggest a much younger age of $3\,\mathrm{Gyr}$. This age is supported by \citet{2024Ne}, whose study focuses on the ages of the metal-rich stars in the Solar neighbourhood.

Looking at other related structures also places limits on the bar's age. It is thought that the nuclear stellar disc formed shortly after the bar (\citet{2020Ba} discuss the usefulness and restrictions of this method for dating the bar), and \citet{2020NL} find that this is dominated by young stars, suggesting that either the bar has been inefficient at inward gas transport for most of its history or else it is just $1\,\mathrm{Gyr}$ old. Favouring the first case, \citet{2023Sa} uses variable stars from the same region to find a minimum age of $8\,\mathrm{Gyr}$, and \citet{2023Sc} suggest that the nuclear disc is even older. These ages are consistent with the minimum age of the inner ring, found by \citet{2022Wy} to be $7\,\mathrm{Gyr}$.. Despite no agreement, evidence is mounting for an older bar, and it is this earlier formation scenario that we examine the mechanisms of in this study.


Also occurring within $\sim2\,\mathrm{Gyr}$ of these older estimates, a significant accretion event occurred $\sim8$-$11\,\mathrm{Gyr}$ ago (this range of times is consistent with the results of \citet{2018Be,2018He,2019Di,2019Ga,2020Ch,2021Na}), known as Gaia-Encaladus/Sausage (GES). Such an accretion event had been hinted at for some time \citep{2002Br,2007Br}, but was confirmed by \citet{2018Be} and \citet{2018He} using data from Gaia \citep{2016Pr}. GES is thought to have had a stellar mass of $10^{8-9.5}\,\mathrm{M}_{\odot}$, around 5-10\% of the Milky Way's at this time \citep[e.g.][]{2018He,2019De,2020Fe,2023La}. The Milky Way is expected to have had a quiescent merger history after this merger event up to the present day \citep{2015Ru,2020Fr}, making GES one of the most significant events in our galaxy's history. While uncertainties in the time of this merger and the time of bar formation are much larger than the expected dynamical times involved, the estimated times of both significantly overlap. Because of this overlap and the potential impact mergers can have on bar formation, there may be some link between the two events.  Moreover, the lack of significant mergers after this event suggests that the any structures formed at this epoch would evolve by secular processes rather than being disrupted by external interactions.

As bars evolve, they have a great impact on the evolution of the galaxies they reside in, through the outwards transfer of angular momentum \citep[e.g.][]{1972Ly,1984Tr,2003At} and inflow of matter \citep{2002At}. As such, the lasting impact of the GES merger on the Milky Way's present day structure is likely to be significant, both as its last significant merger and possible links to the formation of its bar. We use a pre-existing zoom-in cosmological simulation to investigate this link as the focus of this paper, including rerunning the simulation with the GES analogue removed to examine its evolution without the influence of its last significant merger. 

In section \ref{meth} we describe the Au-18 simulation, the numerical parameters we extract from it, and the details of the new approach used for our rerun. We present our results in section \ref{allresults}, and discuss the implications of these results on bar formation in the Milky Way and simulated analogues in section \ref{GESMilk}. Finally, in section \ref{stunt}, we briefly examine the wider effect of the GES merger on our main Milky Way analogue including the extent and orientation of its disc.

\section{Methodology}
\label{meth}

\subsection{The Auriga simulations}
\label{ogmeth}

The Auriga simulations \citep{2017Gr} are a set of cosmological magnetohydrodynamical zoom-in simulations of Milky Way mass halos, comprising 30 halos in the virial mass\footnote{We define the virial mass, $M_{200}$, as the mass contained inside the radius at which the mean enclosed mass density equals $200$ times the critical density of the universe.} range $1-2\times10^{12}\,\mathrm{M}_\odot$ and 10 halos in the lower mass range  of $5-10\times11^{12}\,\mathrm{M}_\odot$. These halos from a $\mathrm{\Lambda}$ cold dark matter only version of EAGLE \citep{2015Sc}, a large-volume cosmological simulation with a $100\,\mathrm{Mpc}$ side-length periodic box, are randomly selected from the most isolated quartile of halos in the corresponding mass range at $z=0$. Auriga necessarily adopts the same cosmology as EAGLE, with $h=0.677$, $\Omega_{\mathrm{b}}=0.048$, $\Omega_{\mathrm{m}}=0.307$, and $\Omega_{\mathrm{\Lambda}}=0.693$. These values are physically motivated by \citet{2014Pl}. While a detailed description of the Auriga simulations can be found in \citet{2017Gr} and references therein, we briefly describe the simulations below. Additionally, the simulations have been made publicly available \citep{2024Gr} and are accessible via \href{https://wwwmpa.mpa-garching.mpg.de/auriga/data.html}{https://wwwmpa.mpa-garching.mpg.de/auriga/data.html}.

For each halo, its final dark matter content within 4 virial radii is traced back to $z=127$, and the particles in the corresponding region are replaced with a large number of lower mass particles which are subsequently split into dark matter and gas cells according to the cosmological baryon mass fraction. This leads to initial conditions with typical resolution in the main halo forming region of $\sim3\times 10^{5}\,\mathrm{M}_{\odot}$ in dark matter and $\sim5\times10^{4}\,\mathrm{M}_{\odot}$ in baryons, with decreasing dark matter particle/gas cell resolution at greater distances from this region. The result is a cosmological zoom-in simulation, with a highly resolved central galaxy within a computationally cheap surrounding Universe that provides the large-scale tidal field of the Cosmic web. From this point, the initial conditions are evolved up to $z=0$ for each halo. The dark matter and stars are given a softening length of $500h^{-1}\,\mathrm{cpc}$ up to $z=1$ and $369\,\mathrm{pc}$ beyond this. The gas cells' softening lengths are allowed to vary from this as a minimum up to a maximum of $1.85\,\mathrm{kpc}$ and scale with the size of the gas cell.

The simulation was run using the magneto-hydrodynamic code \textlcsc{AREPO} \citep{2010Sp,2016Pa} which uses a moving-mesh approach to evolve the gas. The Auriga physics model includes many processes important for galaxy formation \citep[for full details, see][]{VGS13,MPS14,2017Gr}. For brevity, the main ingredients of the model are: primordial and metal line cooling; a uniform UV background that completes at $z=6$ \citep{FG09}; a two-phase subgrid model for the interstellar medium \citep{SH03}; a stochastic star formation model in gas denser than $0.1$ atoms $\rm cm^{-3}$ that spawns a star particle representing a simple stellar population; stellar evolution and feedback from AGB stars and type Ia and type II supernovae; black hole seeding, accretion and AGN feedback; and magnetic fields \citep{PMS14,PGG17}. 

\subsubsection{Auriga 18}

\begin{figure*}
\centering
\includegraphics[width=0.9\textwidth]{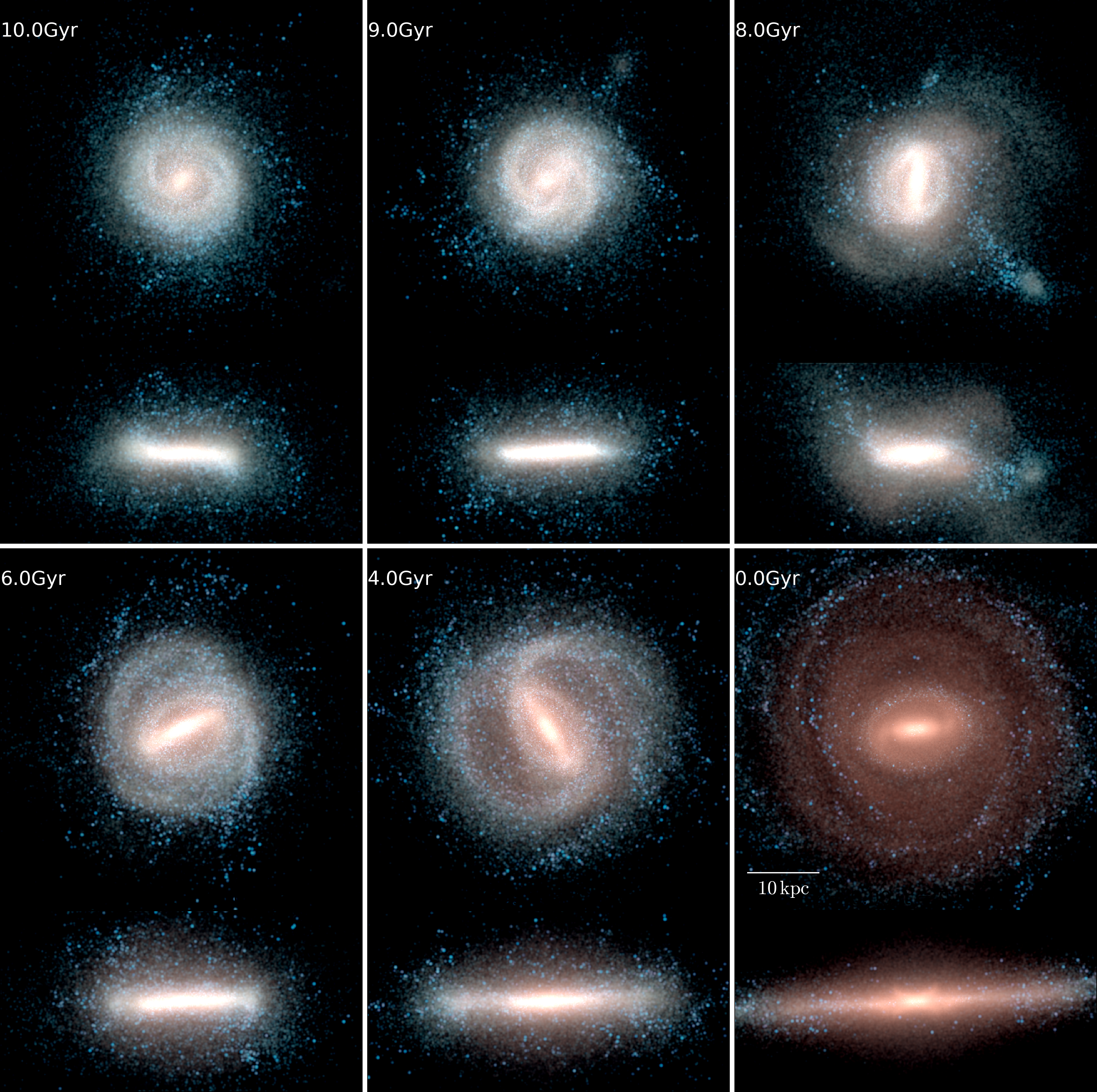}
\caption{Six face-on and edge-on projections of Au-18 at various snapshots throughout the simulation. Each takes the star particles' K-, B- , and U- band luminosities and shows them here in red, green, and blue respectively. This gives the older star particles in red and the younger star particles in blue.}
\label{images}
\end{figure*}

In this paper we focus on the Au-18 simulation, as this galaxy has previously been found to be a close Milky Way analogue. \citet{2020Fr} examines the chemodynamic properties of the discs and bulges of the Auriga galaxies and identifies Auriga 17 and 18 (Au-17 and Au-18) as being very similar to the Milky Way. They find that the metal-poor stars in these galaxies' bulges have a thick disc-like morphology and are rotating almost as quickly as the metal-rich population, consistent with the Milky Way \citep{2013Ne}. In addition, both galaxies have bars and their bulges show a prominent boxy/peanut shape which, for Au-18, can be prominently seen in the edge-on projection at $0.0\,\mathrm{Gyr}$ in Fig. \ref{images}. \citet{2019Fa} also identifies Au-17 and Au-18 (among others) as having a halo dominated by metal-rich, highly eccentric stars similar to the population identified as belonging to GES in \citet{2018Be} and \citet{2018He}. Both of these simulations represent a path of evolution to create the Milky Way as seen today, including its bar, so examining the cause of their bar's formations can shed light onto the Milky Way's real evolution. Of the two, Au-18 exists with more frequent snapshots, so we choose this simulation for our study here.

Au-18's chemodynamics are a result of a merger history which is also similar to that of the Milky Way. Au-18 has a quiescent merger history, with no major mergers in the last $12\,\mathrm{Gyr}$ of its evolution \citep{2020Fr}. Its radially biased metal-rich component in the inner halo is a result of a GES-like merger which Au-18 experiences during this quiescent period at around $9\,\mathrm{Gyr}$ ago \citep{2019Fa} with a lower but broadly consistent stellar mass ratio of $5\%$ \citep{2020Fr}. The latter stages of this merger can be seen at $8.0\,\mathrm{Gyr}$ in Fig. \ref{images}. We will refer to this merging dwarf galaxy in Au-18 as GES-18.

\subsection{Re-simulation}
\label{rerunmeth}

In order to explore the effects caused by the GES-18 merger itself on our Milky Way analogue, we rerun the Au-18 simulation ourselves without GES-18, aiming to modify only this element of the galaxy's history. To achieve this, we use the merger-tree to identify the unique identifiers of each dark matter particle belonging to the GES-18 halo at the snapshot when it is at its peak halo mass. We then remove all of the particles with these identifiers at the earlier snapshot corresponding to $z=4$. The resultant altered version of this snapshot is then used as the initial conditions from which we run the simulation again. We use the same code and outputs as for the original 3000 snapshot Au-18, with the only difference being a later start-time with some particles removed as described.

While we do not remove the baryonic matter associated with GES-18's progenitors, the halo is still in the process of assembling at this time and therefore its baryonic content is too scattered (and too light without the dark matter) to form the galaxy. The removal does also result in a slight change in cosmology due to the marginally decreased dark matter content of the simulation box, as well as a more significant reduction in the final mass of our main halo. The former is a small enough change that we would not expect any significant effects by the end of the simulation, while the latter is to be expected from the removal of a merger. We use the merger tree to identify the halo corresponding to Au-18 in each snapshot.

\citet{2023Re} also alter a simulation to study the impact of GES-like mergers on a Milky Way like host, producing 5 examples of merger histories with GES-like mergers at different mass ratios. They use genetic modification of their zoom-in simulation's initial conditions, giving a consistent background cosmology and final halo mass in each case by assigning the mass removed from their GES-analogue elsewhere. However, for the purposes of our study, we opt for simply removing the GES-18 halo in order to reduce the impact of changing other aspects of Au-18's accretion history, even if it leads to a less massive halo.

The main halo of our rerun is visually indistinguishable from the original simulation up until a lookback time of $9\,\mathrm{Gyr}$, immediately before the GES-like merger event in the original run. The rerun does not experience this event, but it does experience its own later merger at a lookback time of $8\,\mathrm{Gyr}$. This is the remnant of a secondary dwarf galaxy which, in the original simulation, merges with GES-18 after its first pericentre around Au-18. This new merger is retrograde and occurs at a lower mass ratio than in the original, so has less of an effect on the disc. The differences beyond $9\,\mathrm{Gyr}$ lookback time are therefore due to the GES-18's removal, rather than any changes to the cosmology or further changes to the halo's wider accretion history.

\subsection{Calculation of bar properties}
\label{calcs}

At each snapshot (with lookback times $t_{\mathrm{lookback}}$) we re-centre the main halo using its star particles. In three stages, we take a sphere around the origin, calculate the centre of stellar mass for that sphere, and adjust the origin to match this. We carry this out for spheres of radius $30\,\mathrm{kpc}$, $10\,\mathrm{kpc}$ and $4\,\mathrm{kpc}$, ensuring an appropriate centre for the halo as a whole, the disc and the inner bar forming regions respectively. Then, taking the star particles younger than $2\,\mathrm{Gyr}$ and within $10\,\mathrm{kpc}$ of this centre, we align the z-axis with their angular momentum vector so the disc lies in the x-y plane.

Our main diagnostic for bar formation is the bar strength, which we define as the maximum of the $m=2$ Fourier mode of the galaxy surface density \citep[e.g][]{2020Fr}. Using the stellar particles within the plane of the disc, up to a maximum height of $1\,\mathrm{kpc}$, we calculate the contribution of the second Fourier mode ($A_2$) to the distribution of these particles at a set radius $r$ by:

\begin{equation}
    \label{Fourier}
    A_2 = \sqrt{\left( \sum_k m_k \cos{2 \theta_k} \right) ^2 + \left( \sum_k m_k \sin{2 \theta_k} \right) ^2} / \sum_k m_k
\end{equation}

where $m_k$ is the mass and $\theta_k$ is the angle from the vector given by $y=0,x>0$ within the disc plane for each particle $k$ at that radius. We carry this out in $0.25\,\mathrm{kpc}$ wide annuli in the disc centred on the origin out to $30\,\mathrm{kpc}$. We define an edge of the disc ($R_{\mathrm{disc}}$) at each snapshot by the last such ring to contain an average face-on surface density of star particles above $1\,\mathrm{M}_{\odot}\,\mathrm{pc}^{-2}$. At any given time, our bar strength ($A_{2,\mathrm{max}}$) is then the maximum $A_2$ value found in any ring within the disc up to its edge. This measure can be increased by spiral arms outside the bar or ongoing mergers, so we also use the full radial profile of the $A_2$ mode to confirm the presence of a bar.

We also find the bar's length ($R_{\mathrm{bar}}$, strictly the half-length), angular velocity, and corotation radius. Our bar length is defined as the first radius outside the annulus used for the bar strength at which the value of $A_2$ drops below $60\%$ of the bar strength.

For GES-18 we only record global properties: its total mass ($M_{\mathrm{GES}}$), the distance between its centre and the main halo's ($d_{\mathrm{GES}}$), and the tidal field it induces at the centre of the main halo ($g'=\mathrm{G}M_{\mathrm{GES}}d_{\mathrm{GES}}^{-3}$).

\section{Results}
\label{allresults}

\subsection{Bar formation in Au-18}
\label{ogresults}

\begin{figure*}
\centering
\includegraphics[width=0.9\textwidth]{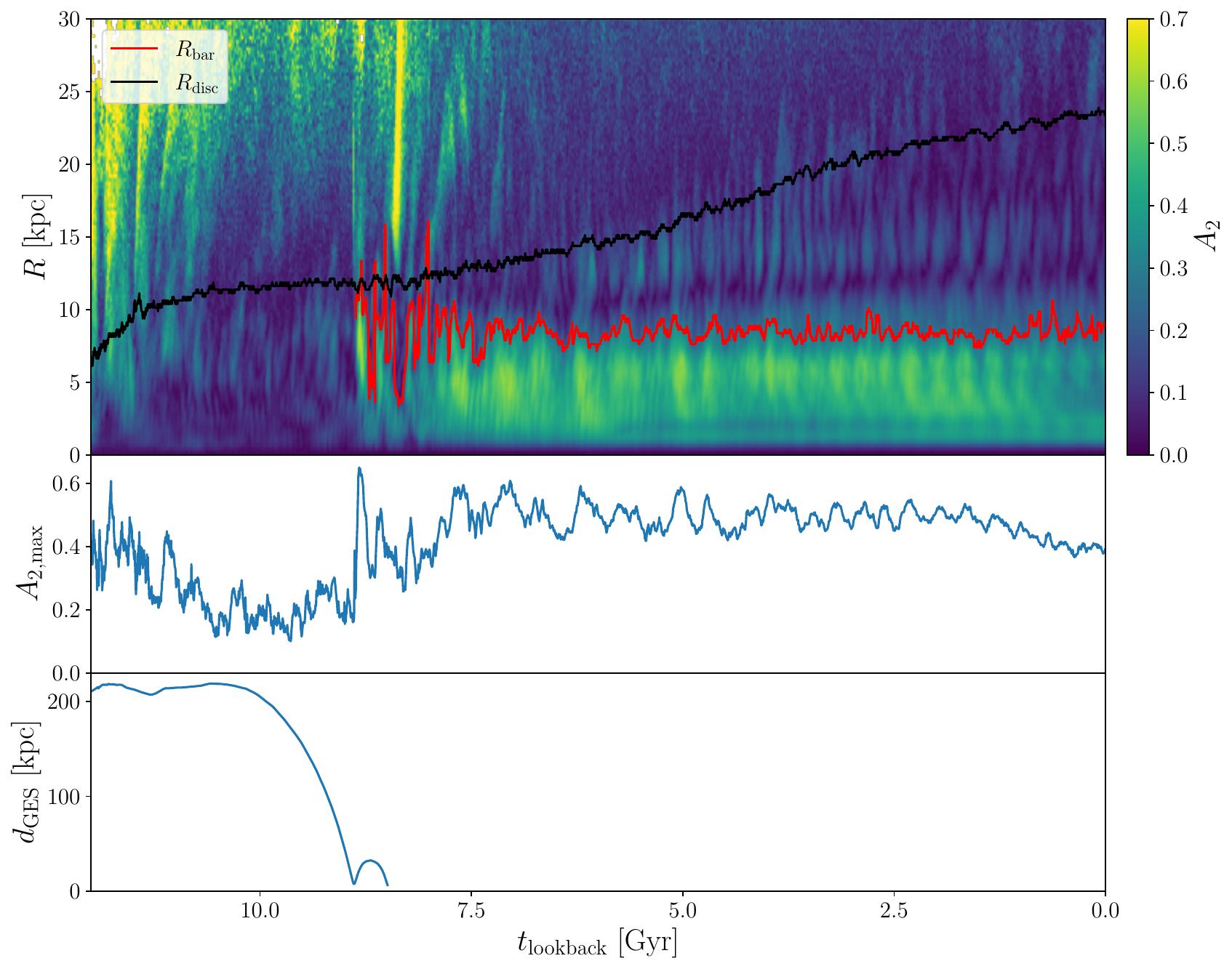}
\caption{Evolution of the bar and the disc throughout the simulation. The colour scale in the upper panel gives the magnitude of the $A_2$ Fourier mode in the stellar disc as a function of lookback time and radius in the disc plane. We overlay the extent of the bar and the disc in red and black respectively, with the bar length only being shown after the bar strength has increased beyond 0.25 for the final time. The middle panel shows the stellar bar strength, as calculated in section \ref{calcs} from the maximum $A_2$ mode within the extent of the disc. The lower panel shows the physical distance between the centres of the main Au-18 halo and GES-18 until the time of their merger.}
\label{fulltime}
\end{figure*}

Fig. \ref{fulltime} shows the evolution of the bar and the GES-18 merger event throughout Au-18's history. The top panel shows a 2D histogram of $A_2$ as a function of radius and time, and the second and thirds panels show the bar strength (maximum $A_2$ within the disc) and distance between GES-18 and Au-18 respectively as functions of time. By examining the bottom panel, we see that GES-18's main progenitor forms $\sim200\,\mathrm{kpc}$ from Au-18 before beginning its approach at a lookback time of around $10\,\mathrm{Gyr}$. This approach reaches its first pericentre at $8.9\,\mathrm{Gyr}$ ago, before splashing back to a distance of $30\,\mathrm{kpc}$ and finally merging with Au-18 at $8.5\,\mathrm{Gyr}$ as defined by the merger-tree code.

The time around the first pericentric passage of GES-18, i.e. between $9-8.5\,\mathrm{Gyr}$ appears to separate two phases of Au-18's evolution. During the $2-3\,\mathrm{Gyr}$ before this, the disc is relatively axisymmetric in its inner regions, as can be seen by the low $A_2$ values in the middle panel of Fig. \ref{fulltime} with the extra radial information of the top panel showing spiral arms and some amount of left-over turbulence from its formation maintaining $A_{2,\mathrm{max}}\approx0.2$. The extent of the disc itself is also relatively constant during this time. After this transition, Au-18 forms a bar which persists until the end of the simulation at a relatively constant strength of $0.5$. The length and corotation radius of this bar are also constant from $7.5\,\mathrm{Gyr}$ onwards ($8\,\mathrm{kpc}$ and $9\,\mathrm{kpc}$ respectively, making for a fast bar), while an increased rate of disc growth is maintained until the end of the simulation so that the bar is a similar size to the disc early on but only exists in the inner third by $z=0$.

\begin{figure*}
\centering
\includegraphics[width=0.9\textwidth]{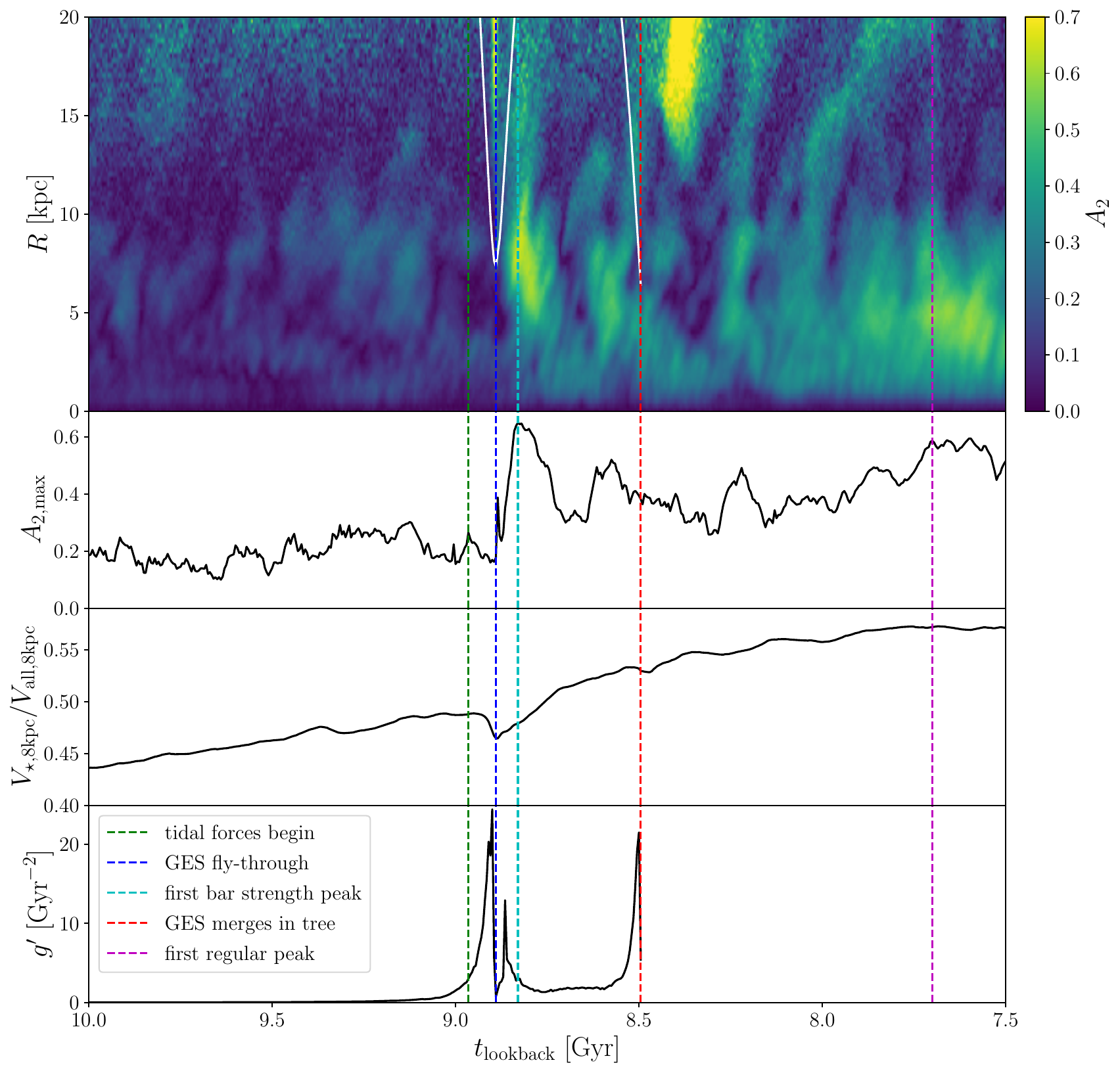}
\caption{The evolution of a selection of galaxy properties in Au-18 at lookback times of $10-7.5\,\mathrm{Gyr}$. Top panel: magnitude of the $A_2$ Fourier mode in the stellar disc at radii up to $20\,\mathrm{kpc}$. The distance from Au-18 to GES-18 is overlaid in white. Second panel: bar strength. Third panel: baryon dominance within the inner $8\,\mathrm{kpc}$ of the disc, shown as the rotation velocity due to stars at $8\,\mathrm{kpc}$ divided by the total rotation velocity at the same radius. Bottom panel: The approximate tidal field induced by GES-18 at the centre of Au-18. The vertical lines indicate the time at which, from left to right: tidal forces first become significant in Au-18 (green); GES-18 experiences its first pericentre around Au-18 (blue); the bar strength first peaks significantly (cyan); GES-18 and Au-18 become one subhalo in the simulation's merger tree (red); the bar strength reaches the oscillating state in which it remains for the next $5\,\mathrm{Gyr}$ (magenta).}
\label{zoomtime}
\end{figure*}

Fig. \ref{zoomtime} provides a closer look at how Au-18 changes during this transitional epoch. In the upper panel we show a portion of Fig. \ref{fulltime}'s upper panel to zoom in on the radial dependence of $A_2$ for lookback times $10-7.5\,\mathrm{Gyr}$, with the distance between GES-18 and the centre of the main halo represented by the white curve. The other three panels show the bar strength, inner baryon dominance, and tidal field strength across the same time period. Prior to GES-18's first pericentre, the $A_2$ mode is most prevalent at radii of $5-10\,\mathrm{kpc}$ although still weak. This is indicative of spiral arms in the outer disc, features which can be seen in the upper-left and upper-middle panels of Fig. \ref{images}. There are low amplitude perturbations in the inner disc, but these are weaker than $0.2$, the value typically used to represent the formation of a bar.

Immediately after the first pericentre at $8.9\,\mathrm{Gyr}$, the $A_2$ spikes in the outer disc. This perturbation moves inwards within $100\,\mathrm{Myr}$ and settles across the inner $4\,\mathrm{kpc}$ for $200\,\mathrm{Myr}$ before dissolving into axisymmetry. At the same time, a second perturbation is formed in the outer disc and similarly moves inwards so that there is again a bar in the inner $4\,\mathrm{kpc}$ just $100\,\mathrm{Myr}$ after the first bar dissolved. It is worth noting that this second perturbation does not follow GES-18's second pericentre, in fact preceding it by $200\,\mathrm{Myr}$. This pattern of dissolving and being replaced by a new perturbation from the outer disc occurs a third time before the inner $6\,\mathrm{kpc}$ form a long-lasting bar, whose length and strength settle towards $8\,\mathrm{kpc}$ and $0.6$ respectively by $7.7\,\mathrm{Gyr}$. The pattern described matches that occurring in \citet{2018Lo} where, in multiple cases, a purely tidal interaction forms a bar by first perturbing the less gravitationally bound outer regions. We examine the implications of this similarity in section \ref{tides}.

Visible in Fig. \ref{fulltime}, there is some oscillation continuing to $z=0$, with the bar strength oscillating between $0.45$ and $0.55$ and the bar length between $7.5-10\,\mathrm{kpc}$, both with a period of $\sim0.25\,\mathrm{Gyr}$. As \citet{2020Hi} shows, this can be caused by the bar interacting with the spiral arms, since as spiral arms line up with the ends of the bar they appear to lengthen the bar and increase the $A_2$ mode. The oscillation may also be a projection effect due to our alignment of the galaxy, since this period almost exactly matches the bar's pattern speed of $\sim28\,\mathrm{rad}\,\mathrm{Gyr}^{-1}$.

The second panel of Fig. \ref{zoomtime} shows the more widely used, single-valued, bar strength measurement of $A_{2,\mathrm{max}}$, as described in section \ref{calcs}. This provides broadly similar information of a weak/non-existent bar up until a sudden jump at $8.8\,\mathrm{Gyr}$, followed by a period of wide variation at lower amplitudes, and ending with a higher, more consistent bar strength up until a buckling event from $1.5\,\mathrm{Gyr}$ onwards slightly reduces the bar strength. In this case though, this measure is limited in details. Au-18's spiral arms contribute to a higher early strength, and the outer disc perturbations are indistinguishable from a strong bar. While we continue to use this measure for a quantifiable bar strength, we use the $A_2$ radial profile to help identify the features which contribute to the strength.

Au-18's stellar profile is also changed around this time. The third panel, even accounting for the halo-mixing effects (see section 7.4 of \citet{2021Sp} for a description of this effect) as GES-18 infalls, shows a sharp increase in the baryon dominance inside $8\,\mathrm{kpc}$ ($V_{\star,8\mathrm{kpc}}/V_{\mathrm{all},8\mathrm{kpc}}$) from $8.9-8.5\,\mathrm{Gyr}$. This is the greatest departure from Au-18's normal evolution of baryon dominance across the entire simulation, and implies either a large depositing of stars directly from GES-18, a sudden increase in star formation or a transition to a more centrally concentrated stellar density profile as soon as GES-18 first interacts with Au-18. We examine this further using Fig. \ref{baryonmovement} below.

The bottom panel of Fig. \ref{zoomtime} shows the tidal field placed on the main halo by GES-18. This is negligible for most of the time shown, but peaks sharply at both GES-18's first pericentre and its merger time. The first of these peaks in tidal forces is the time at which many of the previously discussed changes in Au-18 occur.

\begin{figure}
\centering
\includegraphics[width=\columnwidth]{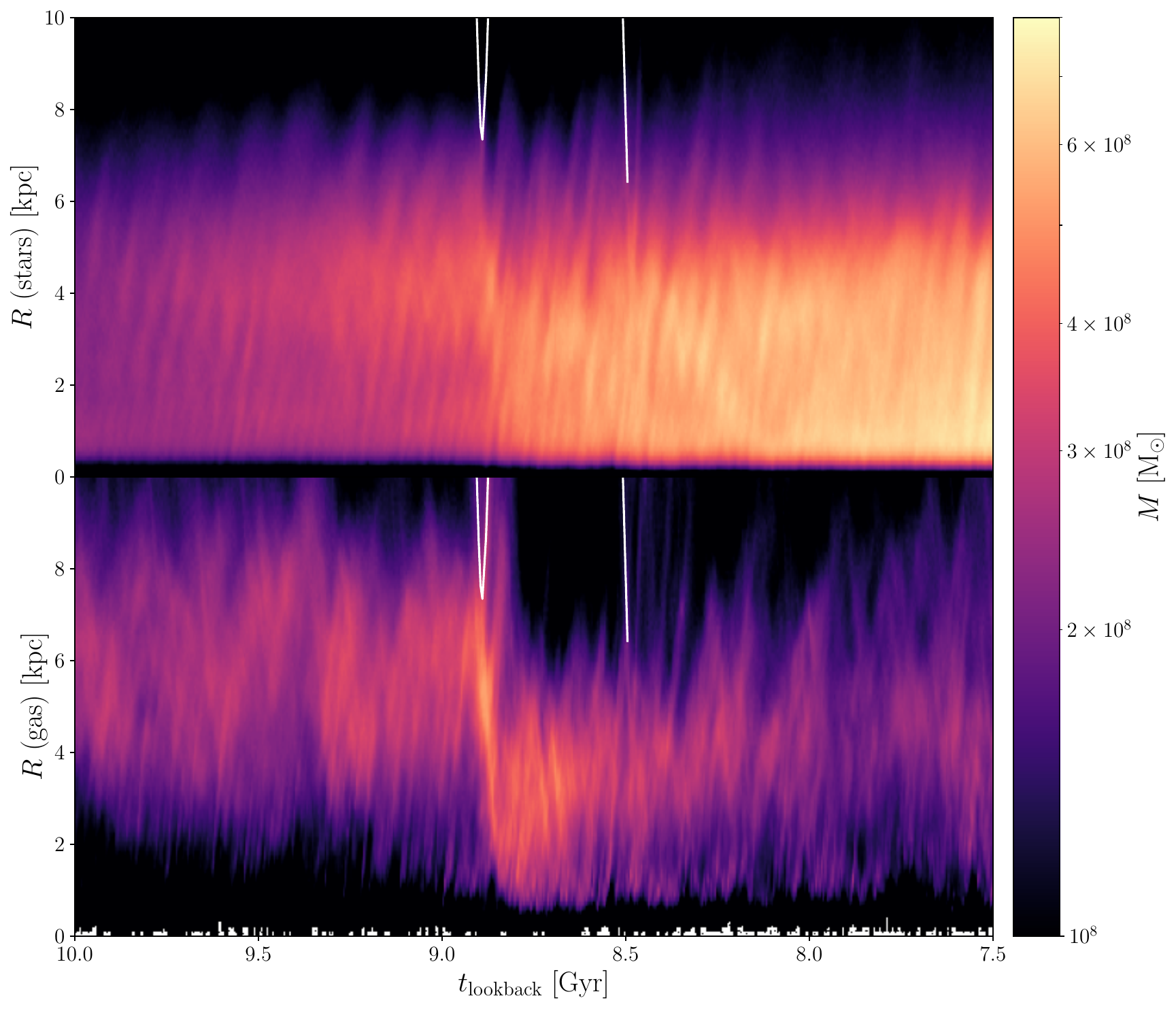}
\caption{The total mass at each snapshot in each $0.1\,\mathrm{kpc}$ annulus of star particles (upper) and gas cells (lower) in Au-18 from lookback time of $10-7.5\,\mathrm{Gyr}$. The colour scale showing this mass is logarithmic. In both panels, we show the distance between the centres of GES-18 and the main halo by the white line.}
\label{baryonmovement}
\end{figure}

Fig. \ref{baryonmovement} shows the radial distributions of stars (top) and gas (bottom) over the same time period as in Fig. \ref{zoomtime}, and suggests a combination of increased star formation and a shift in stellar density profile occurring directly after GES-18's first pericentre. The lower panel shows a galaxy-wide migration of gas in Au-18 beginning $8.9\,\mathrm{Gyr}$ ago, with the mass distribution's peak shifting from a radius of $6\,\mathrm{kpc}$ to $4\,\mathrm{kpc}$ as a result. Furthered by accretion of gas from GES-18, this creates a very dense environment of gas in the inner few $\mathrm{kpc}$ of the disc. In the upper panel, we see a less extreme version of this migration happen for the stellar component of Au-18, visible as a reduced stellar mass in the $5-6\,\mathrm{kpc}$ region. However, the dominant change in stellar distribution at GES-18's first pericentre is a sudden increase in star formation in the inner $4\,\mathrm{kpc}$ of the disc. This star-burst is sustained for almost $1\,\mathrm{Gyr}$, as seen by the depletion of gas in the lower panel and noted in \citet{2020Gr2}, and fundamentally changes the stellar distribution in Au-18. Therefore the migration of gas inwards following GES-18's first interaction with Au-18 appears to induce a dramatic rise in the dominance of stars in Au-18's inner disc.

\subsection{Comparison to a controlled counterpart simulation}
\label{rerunresults}

\begin{figure*}
\centering
\includegraphics[width=0.9\textwidth]{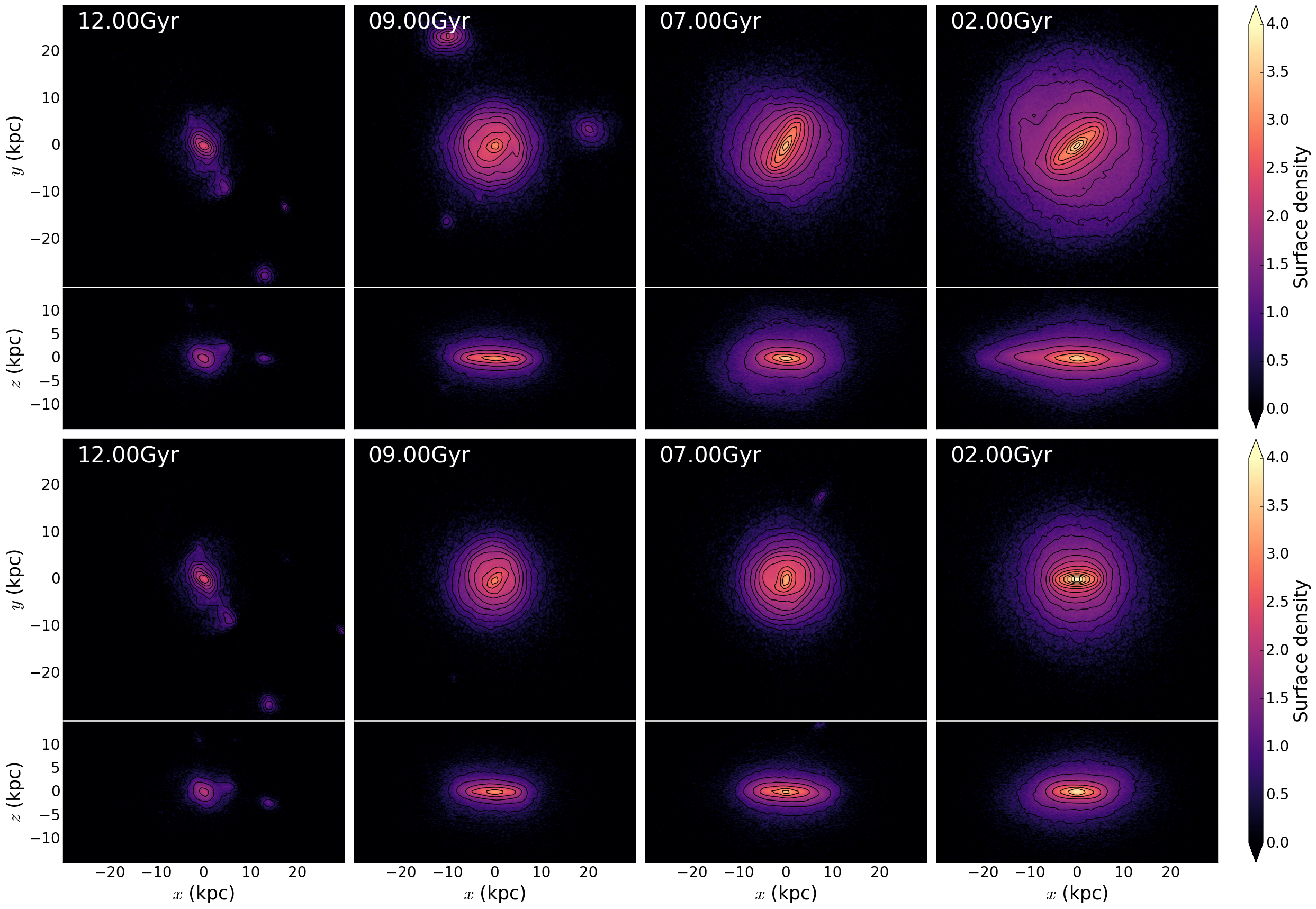}
\caption{Face-on and edge-on stellar surface density profiles for each of Au-18 (top) and our rerun (bottom) at four different snapshots: $12\,\mathrm{Gyr}$ shortly after the initial conditions of the rerun; $9\,\mathrm{Gyr}$ as GES-18 approaches in the original; $7\,\mathrm{Gyr}$ following the interaction period; $2\,\mathrm{Gyr}$ once both simulations have almost fully evolved but before the rerun's polar disc dominates the young stellar population. The surface density colour-scale is measured as the particle count per bin and is logarithmic.}
\label{sidebyside}
\end{figure*}

In this section we present results from our controlled resimulation described in section \ref{rerunmeth}, in which we remove the GES-like galaxy from the simulation before it enters the main halo of our Milky Way like galaxy. In the left panels of Fig. \ref{sidebyside}, we confirm that the main halos of each simulation match up until GES-18's first interaction in the original simulation. However, by a lookback time of $7\,\mathrm{Gyr}$, shown in the middle panels, only Au-18 has a bar. In addition, the rerun's disc is less extended. By close to the end of the simulations, shown at $2\,\mathrm{Gyr}$ in the right panels, this difference in disc extension has grown dramatically and the rerun now has a denser central stellar profile. More crucial to this study's question, the rerun also has a bar by this time; the removal of GES-18 has delayed bar formation rather than stopped it.

\begin{figure}
\centering
\includegraphics[width=\columnwidth]{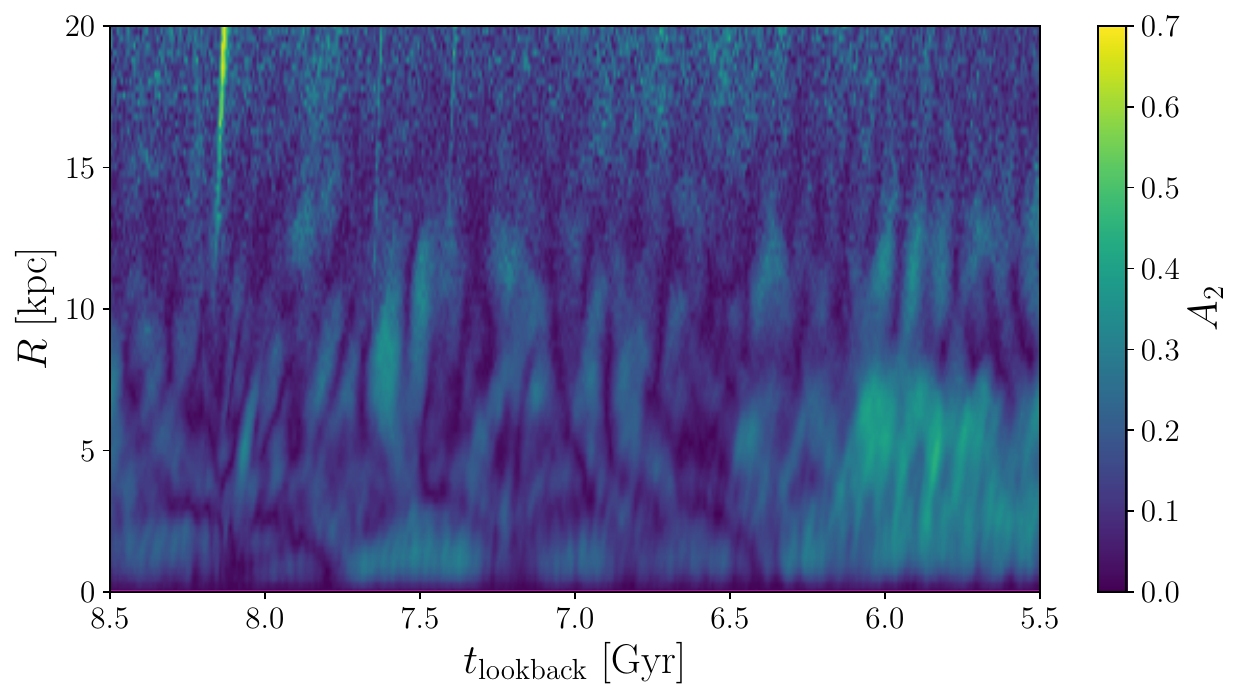}
\caption{The magnitude of the $A_2$ Fourier mode in the stellar disc of our re-simulation, at radii up to $20\,\mathrm{kpc}$ and lookback times from $8.5-5.5\,\mathrm{Gyr}$. The short peak in the outer regions shows the first pericentre time of the smaller merger. Note the absence of the tidal features clearly visible in Fig. \ref{zoomtime}.}
\label{newzoom}
\end{figure}

The rerun experiences another somewhat significant merger even with the removal of GES-18. The first pericentre is visible in Fig. \ref{newzoom} as a transient peak in the $A_2$ mode outside of $10\,\mathrm{kpc}$. This occurs at the later lookback time of $8.2\,\mathrm{Gyr}$ and at $\sim25\%$ of the total mass ratio and $\sim40\%$ of the stellar mass ratio of GES-18 in the original. As outlined in section \ref{rerunmeth}, this dwarf galaxy is involved in the original merger event, but only merges with GES-18 between GES-18's first and second pericentre, and as such was not removed. Without the influence of GES-18, this dwarf galaxy takes longer to reach the main halo, but does eventually merge in the rerun.

\begin{figure*}
\centering
\includegraphics[width=0.9\textwidth]{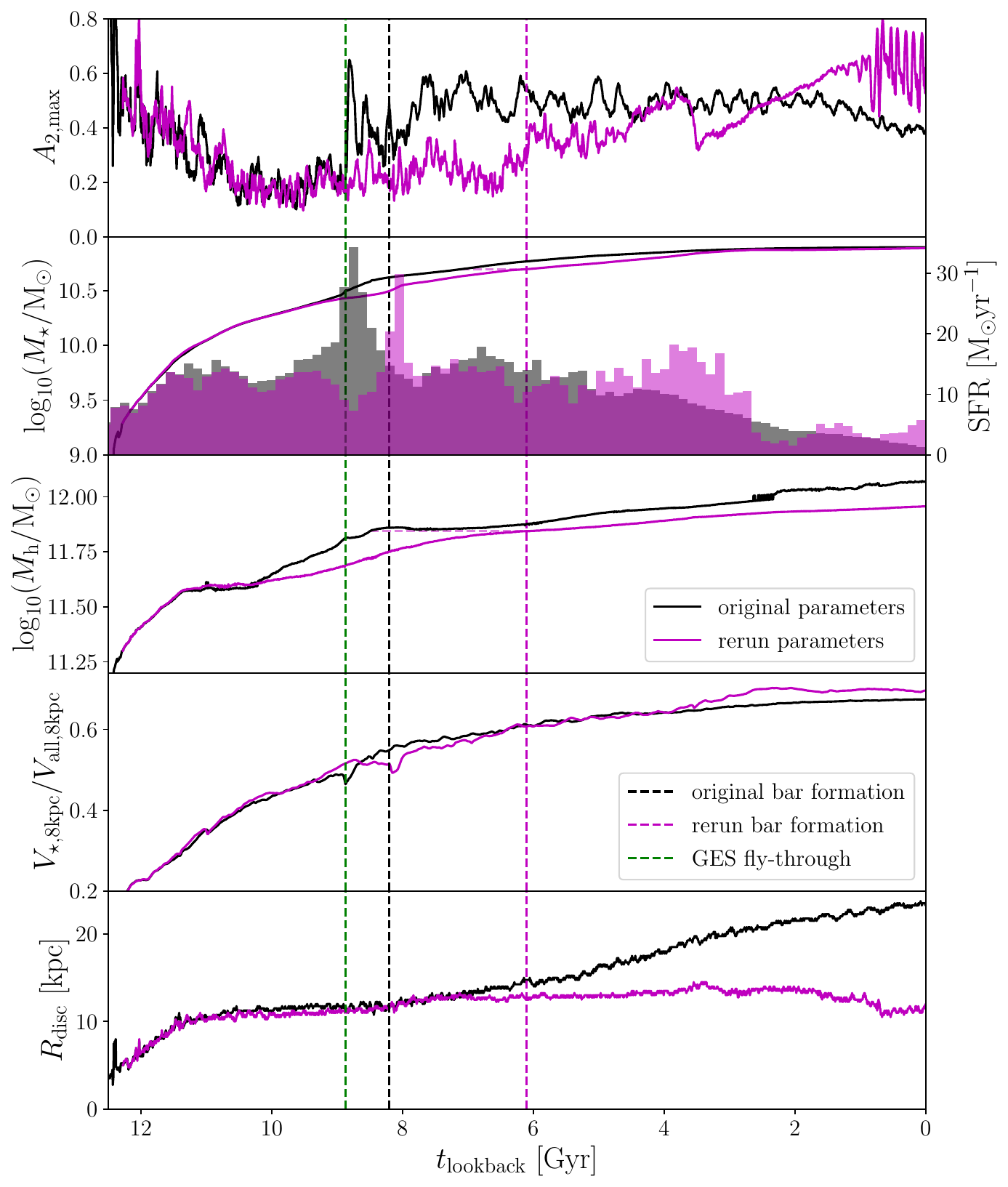}
\caption{Comparison between Au-18 (black lines) and our rerun (magenta lines) of a selection of galaxy properties from the rerun's beginning until $z=0$. Top panel: bar strength, from the $A_2$ Fourier mode. Second panel: total stellar mass. Third panel: total mass of the dark matter halo. Fourth panel: baryon dominance within the inner $8\,\mathrm{kpc}$ of the disc, shown as the rotation velocity due to stars at $8\,\mathrm{kpc}$ divided by the total rotation velocity at the same radius. Bottom panel: radial extent of the stellar disc. The vertical, dashed lines show, from left to right, the time of the GES-18's first pericentre in the original, Au-18's bar formation time (by visual inspection of the full $A_2$ profile), and our rerun's bar formation time. We also include horizontal lines to indicate the time at which Au-18 had the same properties as our rerun at the time of the rerun's bar formation.}
\label{newsum}
\end{figure*}

Despite this merger, the rerun does not form a bar at $8.2\,\mathrm{Gyr}$ ago either. Fig. \ref{newzoom} and Fig. \ref{newsum} shows that while there is an excitation of the $A_2$ mode in the inner regions of the disc following the new merger, it only reaches a maximum amplitude of $0.3$ before fading back to its state before the merger. This lasts $1.5\,\mathrm{Gyr}$ and at no point during this time does it form a visually distinguishable bar, instead having strong spiral arms which occasionally converge in the galaxy's centre.

The rerun's bar, which lasts until $z=0$, in fact forms not long after the disappearance of this perturbation, growing from $6.4-6.0\,\mathrm{Gyr}$. The bar experiences $1.5\,\mathrm{Gyr}$ of stability at $A_{2,\mathrm{max}}=0.4$ before growing further and buckling. During this period of stability it is therefore weaker than Au-18's bar. In addition, the $A_2$ radial profile differs from the original's during bar formation. Despite strong spiral arms producing higher $A_2$ in the mid-outer disc, the bar itself appears to quickly grow from the centre of the disc and outwards, rather than from perturbations moving inwards. This lengthening is not a persistent feature, since the bar which results after this growth period in the rerun does have a very similar bar length to the original despite the different radial $A_2$ patterns at their respective formation times. Nevertheless, these differing patterns during the bars' growth suggest a different mechanism of bar formation in the rerun in addition to the simple lack of mergers at the the time of bar formation.

Fig. \ref{newsum} shows properties as a function of time from $12.5-0\,\mathrm{Gyr}$ for both the original and rerun simulations. The top panel compares the simulations' bar strengths. The second panel shows both stellar mass and star formation rates for each, with the dark matter halos' masses in the panel below. The fourth panel shows the inner baryon dominance of each, and the final panel shows the extent of their respective discs.

In the top panel of Fig. \ref{newsum} we clearly see a consistent general behaviour of our bar strength measurement between the two simulations up until GES-18's first pericentre at a lookback time of $8.9\,\mathrm{Gyr}$. At this time, the original's bar strength jumps up dramatically, due to the large perturbations of the outer disc found in section \ref{ogresults}, while the rerun's bar strength continues at a low value. While the original forms its bar, the rerun experiences its own merger but does not form a persistent bar following this. Instead, bar formation occurs at a lookback time of $6.1\,\mathrm{Gyr}$, over $2\,\mathrm{Gyr}$ later than in the original with GES-18. The two galaxies can be seen to have very different evolution at late times, but the important takeaway is that both bars persist to the ends of their simulations.

In the three panels below this, we show the stellar mass, halo mass and previously defined baryon dominance respectively, for both the original and the rerun. These quantities are very similar for both up until lookback time $10\,\mathrm{Gyr}$, with only the halo mass diverging at this point before the accretion of GES-18 in the original. Since inner baryon dominance is not affected by this, the dark matter accretion must occur at larger radii during this Gyr. After GES-18 causes a higher star formation rate in the original, we also see a slowing of dark matter accretion for the next $3\,\mathrm{Gyr}$ before both simulations continue to gain mass, both dark matter and stellar, at similar rates until at least $4\,\mathrm{Gyr}$ before the simulations' ends. Looking at just the times of bar formation in each, we see that when the rerun forms its bar it has a higher stellar mass but a very similar halo mass to the original at its own time of bar formation. This further translates to baryon dominance being much higher at bar formation in the rerun than in the original. The original forms its bar when it attains a baryon dominance of $0.55$, whereas the rerun forms its bar about $2\,\mathrm{Gyr}$ later for a higher baryon dominance of 0.6. This suggests very different internal environments in the two galaxies when they form their respective bars.

We also see a dramatic difference between the galaxies at late times, with the most obvious change being the lack of growth in the extent of the rerun's disc from the bottom panel, when removing GES-18. The rerun also develops a polar disc in the final $4\,\mathrm{Gyr}$ of the simulation, presenting as a reduced disc extent, increased inner baryon dominance, and widely varying bar strength in Fig. \ref{newsum}. We discuss these changes and their causes in section \ref{stunt}, but these diverging properties occur well after both bars are established and do not therefore alter our main results.

\section{Discussion}

\subsection{The link between GES and the Milky Way's bar}
\label{GESMilk}

With the rerun not forming a bar until over $2\,\mathrm{Gyr}$ after the original, we can conclude that our removal of GES-18 is responsible for delaying bar formation. While there is some small difference between the two simulations before a lookback time of $9\,\mathrm{Gyr}$, most notably in the total mass of the dark matter halo in the preceding $1.5\,\mathrm{Gyr}$ as shown in Fig. \ref{newsum}, the inner regions and the stellar component (see Fig. \ref{sidebyside} also) remain practically identical until this time. In particular, our bar strength measurement shows the same behaviour in both simulations exactly until the merger time. Therefore we argue that the delay in bar formation is caused by the removal of the merger event itself, rather than any other unintended changes to the galaxy's accretion history leading up to this point.

In addition, the temporary $A_2$ excitation in the rerun due to its smaller merger at $8.2\,\mathrm{Gyr}$ is an indication that Au-18 would be relatively stable against bar formation for a time without GES-18's merger. Since such an $A_2$ asymmetry dissipates on its own during this time, it is unlikely that internal mechanisms would have been able to form a bar before $6.5\,\mathrm{Gyr}$ ago as the beginnings of such a bar would dissipate in the same way as the observed $A_2$ feature. The removal of GES-18 not only delays bar formation but leaves a galaxy which is stable against bar formation for the next $1.5\,\mathrm{Gyr}$. This in turn strongly suggests that GES-18 triggered the formation of Au-18's bar, particularly when considering the lack of internal bar development in both simulations prior to GES-18's first pericentre.

The mechanism of GES-18's triggering of Au-18's bar formation is less apparent, with the merger being a relatively chaotic time in the galaxy's history and as such many conditions changing quickly. In the following, we will focus on the tidal effects of GES-18 on Au-18 and the change in central baryon dominance immediately following the merger.

\subsubsection{Tidal bar formation}
\label{tides}

The bottom panel of Fig. \ref{zoomtime} shows the approximate tidal field (the spatial gradient of the gravitational field along the line connecting the two bodies' centres of mass) exerted on the centre of Au-18 by GES-18 around the the time of its merger. This sharply reaches its maximum of $24\,\mathrm{Gyr}^{-2}$ at its first pericentre, before just as sharply dropping while GES-18 splashes back and around Au-18. The tidal field peaks again to a similar magnitude at the merger time, when GES-18 experiences its second pericentre. Note that artificial halo-mixing due to overlapping dark matter halos during its closest approach is responsible for the brief decrease to almost $0\,\mathrm{Gyr}^{-2}$ at $8.9\,\mathrm{Gyr}$ as the majority of GES-18's halo is temporarily assigned to the main Au-18 halo instead in the simulation's code.

This initial peak in the tidal field at the first pericentre is also when the large $A_2$ perturbation begins in the outer disc. As described in section \ref{ogresults}, this perturbation soon propagates inwards before disappearing and propagating inwards from the outer regions a few more times. \citet{2018Lo} describes this same pattern of bar formation resulting from a purely tidal interaction. They use a set of N-body simulations to model four fly-bys of identical Milky Way mass galaxies at varying strengths of tidal interaction. The result in each case but the weakest (as determined by the Elmegreen parameter described in equation \ref{Elme} below) is the formation of a bar in the prograde oriented galaxy, which is otherwise stable against bar formation for $3\,\mathrm{Gyr}$. All of these tidal bars are formed via the process of repeated inwards-moving perturbations from the outer disc, similar to what we observe in Au-18, creating multiple instances of bars dissolving before a more permanent bar emerges. Since we see Au-18's bar forming in the same way as these tidal bars, with the bar forming $0.7\,\mathrm{Gyr}$ after this process begins, this suggests that Au18's bar is a result of tidal forces from GES-18's first pericentre.

This is also consistent with the strength of the tidal interaction, as determined by the Elmegreen parameter $S$:

\begin{equation}
    \label{Elme}
    S = \frac{M_\mathrm{p}}{M_\mathrm{g}} \left( \frac{R_\mathrm{g}}{R_\mathrm{p}} \right)^3 \frac{T_\mathrm{p}}{T_\mathrm{g}}
\end{equation}

where $M_\mathrm{p}$ is the total mass of the perturbing galaxy, $M_\mathrm{g}$ is the total mass of the host galaxy, $R_\mathrm{p}$ is the distance between the two galaxies' centres at pericentre, $R_\mathrm{g}$ is the radius of the host galaxy's disc, $T_\mathrm{p}$ is the time taken for the perturbing galaxy to orbit 1 radian around the host, and $T_\mathrm{g}$ is the time taken for stars at the edge of the host's disc to complete 1 radian of their own orbit. The Elmegreen parameter expects a permanent bar to be formed for $S>0.04$ \citep{1991El}. The GES-18 interaction has $S\sim0.09$, which meets this criterion and is also similar to one of the bar-forming cases in \citet{2018Lo}, at $S=0.07$. This parameter encodes most of the broad information on the impact of a tidal interaction, but the angle of the perturber's approach compared to the angle of the galaxy's rotation is also important. Prograde encounters have a much stronger effect than retrograde encounters \citep{2018Lo} and intermediate angles fall on a corresponding scale \citep{2014Lo}. GES-18 approaches at an angle of $\sim70^{\circ}$, which is slightly prograde, but highly inclined. As such, the true tidal effect of the merger is likely to be less than indicated by its Elmegreen parameter. It is worth noting, however, that GES-18 reorients the entire disc with its merger (see section \ref{stunt}) which may reduce the weakening caused by its highly inclined approach.

It is also notable that the rerun does not form its bar in this way. This points to some difference in mechanism between the formation of the two bars, and the lack of significant mergers or fly-bys in the rerun at this time suggests that this difference can be accounted for by a lack of tidal interactions. The somewhat significant merger experienced by the rerun $\sim2\,\mathrm{Gyr}$ before bar formation exerts a maximum tidal field of just $4.7\,\mathrm{Gyr}^{-2}$, 5 times less than GES-18 in the original. This suggests that only Au-18's bar, and not the rerun's, is formed by tidal forces and that the delay in bar formation which the rerun experiences is due to the lower magnitude of tidal interaction being insufficient to trigger bar formation.

\subsubsection{Accelerated internal bar formation}

In section \ref{ogresults} we highlight the change in stellar distribution occurring at and after GES-18's first pericentre, which Fig. \ref{baryonmovement} shows is driven by the gas migrating to the inner regions and subsequently triggering an accelerated star formation rate across the next $1\,\mathrm{Gyr}$. In the second panel of Fig. \ref{newsum}, we confirm that this is indeed an increase in star formation rate rather than purely accretion of stars from GES-18, since the star formation rate has its largest peak at this time, a peak which is also prolonged enough to produce such a dramatic shift in stellar density profile.

Since a high density of stars in the inner regions of a galaxy is conducive to bar formation, as is a low density of gas, this could instead be the trigger for Au-18's bar. However it seems more likely that this effect occurred alongside a tidal bar formation mechanism. The migration of the stars and gas occurs at the same time as the first major $A_2$ peak. The bulk of the migrating matter also follows the radius of the $A_2$ excitation as it moves inwards, suggesting that they occur as part of the same tidal event caused by GES-18. As the iterative outside-in $A_2$ perturbations continue, stars continue to form in the central regions as part of the star formation peak the migration triggers, while the gas density gradually decreases. At a lookback time of $\sim8\,\mathrm{Gyr}$, $V_{\mathrm{\star,8kpc}}/V_{\mathrm{all,8kpc}}$ reaches an approximately stable value and the stellar and gas density profiles appear to stabilise too. This also coincides with the time at which the bar stops oscillating with each new perturbation and becomes permanent. These observations suggest that the tidal forces from GES-18 create the initial bar-seeding perturbations in Au-18 and simultaneously cause the baryon migration inwards. As both processes continue, conditions in the galaxy, specifically its inner stellar and gas densities, become able to support a bar. The continuing perturbations are therefore able to then take hold and form a long-lived and stable bar.

Our rerun confirms that the tidal effects are needed to form a bar at this time in addition to the baryon density. Despite the $2\,\mathrm{Gyr}$ delay in bar formation, the rerun's inner baryon dominance in the third panel of Fig. \ref{newsum} is similar to Au-18's at most times. This means that, as noted in section \ref{rerunresults}, by the time of the rerun's bar formation its inner baryon dominance has increased to be higher than Au-18's was when it formed its own bar. In addition, the rerun's smaller merger, occurring just $0.5\,\mathrm{Gyr}$ after GES-18's in the original, does not form a stable bar despite a similar inner baryon dominance. The obvious difference between the two scenarios is the difference in the magnitude of the merger, with the rerun's being only $25\%$ as massive, suggesting that Au-18 did indeed require the increased tidal forces of a more significant GES-like merger to form its bar at this earlier time.

\subsubsection{The wider Auriga simulations}

\begin{figure}
\centering
\includegraphics[width=0.95\columnwidth]{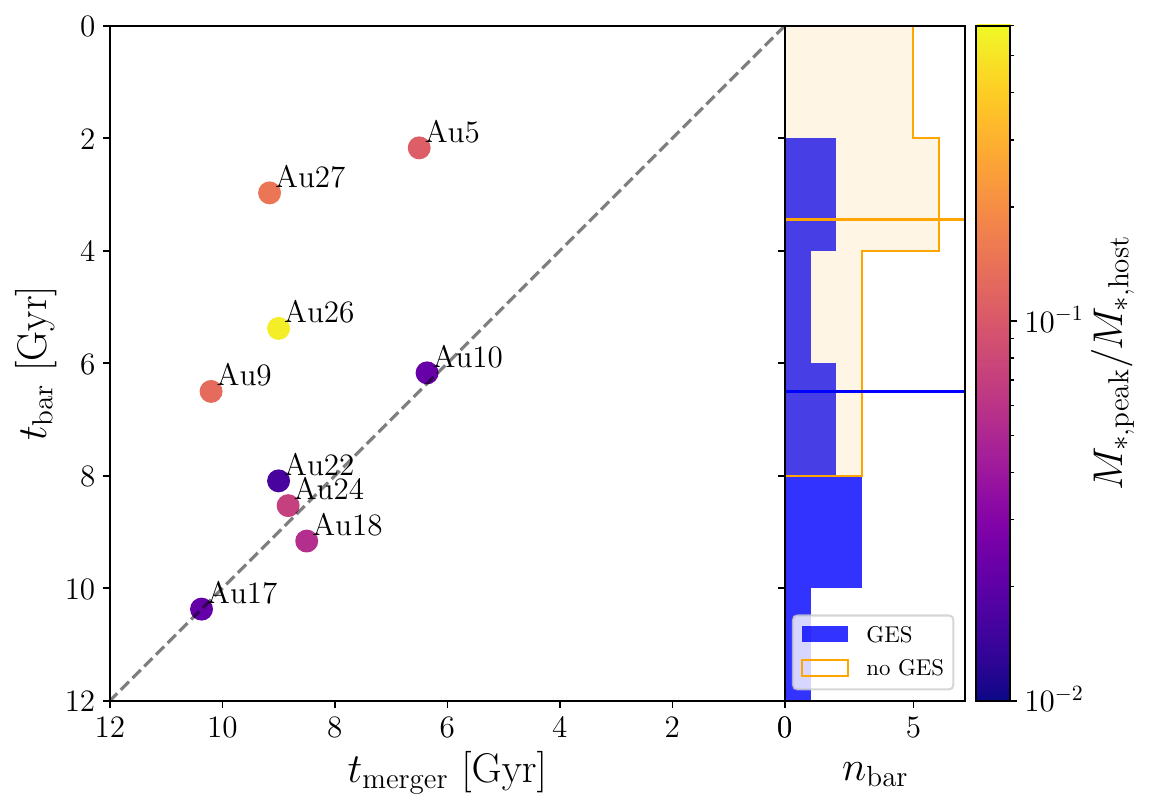}
\caption{The relation between merger times (lookback times on the x-axis from \citet{2019Fa}) and bar formation times (lookback times on the y-axis from Fragkoudi et al. (in prep)) in the Auriga simulations which experience a GES-like merger. Bar formation is defined in this paper as the last time the bar strength passes $A_2=0.25$, a threshold at which bars typically become visually identifiable. We show equal times by the dotted grey line. Points are coloured by the peak stellar mass ratio of the merging galaxy to the main halo \citep{2020Gr2}, with the most relatively massive mergers in yellow. Au-15 is not shown since, although it has a GES-like merger at a lookback time of $5.22\,\mathrm{Gyr}$ at a stellar mass ratio of $\sim0.15$, it does not form a bar. In the right panel, we show histograms of the bar formation times for those Auriga halos with GES-like mergers and those without, in blue and orange respectively. For each histogram we also show the median bar formation time with a horizontal line.}
\label{otherau}
\end{figure}

In this section, we examine the connection between GES-like mergers and bar formation in the context of the larger sample of Auriga simulations. In particular, we highlight the subsample of simulations which \citet{2019Fa} identified to contain a population of radially biased stars in their stellar halos at $z=0$. These populations are found using a Gaussian decomposition of the azimuthal vs radial velocity space of the stellar halo, where halos are determined to be Milky Way-like if the largest component of this decomposition has anisotropy parameter $\beta>0.8$. This mimics the signature of GES which we observe in the Milky Way at present times. In Fig. \ref{otherau} we plot the bar formation times against the merger times of the 9 out of 10 such selected galaxies which form bars before $z=0$, including Au-18. We also show histograms of the bar formation times of galaxies with/without GES analogues in the right panel. We see that the galaxies selected loosely fall into two groups: half of the 10 galaxies form bars within $1\,\mathrm{Gyr}$ of their GES-like mergers, and the other half (including Au-15 which does not form a bar before $z=0$) do not form bars until over $3\,\mathrm{Gyr}$ after their own mergers. No galaxies with GES-like mergers in the sample form bars before the GES merger time (except for Au-18 which, as we have already shown, still forms its bar after GES-18's first close pericentre). In addition, the five galaxies that form bars at the same time as their mergers are the galaxies with the lowest stellar mass ratios between their GES-analogue at its peak stellar mass and the main galaxy at this same time (see the colour bar of Fig. \ref{otherau}). The transition between the two cases occurs at a stellar mass ratio of 10\%, leading to a requirement for GES's stellar mass ratio of less than 10\% in order to create a bar at a time consistent with estimates from e.g. \citet{2020Gr,2022Wy,2023Sa}. This mass is consistent with a stellar mass ratio of 6\% as found by \citet{2018He} and matched by the $5-10\times10^8\mathrm{M}_{\odot}$ stellar mass estimate of \citet{2019De}. \citet{2020Fe} find a higher stellar mass of $10^{8.85-9.85}\mathrm{M}_{\odot}$ giving a ratio of 7-17\%, whereas \citet{2023La} find a lower mass of $1.45\times10^8\mathrm{M}_{\odot}$, implying a 1-2\% ratio (these ratios are under the same assumption for the Milky Way's mass as in \citet{2018He}). Our restriction on the mass ratio required to form a concurrent bar is consistent with even the extremities of stellar masses found for GES. \citet{2020Fr} also gives an upper limit on GES' stellar mass ratio regardless of the bar formation time, requiring <5\% in order to reproduce the rotational velocities of metal-poor stars observed from GES in the Milky Way. This paper is likely to have similar biases to ours due to using the same simulation suite, but both are supported by the relatively low masses from observations.


The Auriga suite as a whole therefore suggests a scenario of bar formation in galaxies with relatively light GES-like mergers. Prior to merging, the galaxy does not contain a bar. When the merger occurs, if the satellite is not too massive, then it induces bar formation in the main galaxy and this bar persists to $z=0$. If the stellar mass ratio is in fact greater than $\sim0.1$, then no bar forms for the next few $\mathrm{Gyr}$, possibly due to the greater disruption to the disc caused by a more massive merger. Importantly, Au-17 and Au-18 fall into the category of smaller mergers which form bars almost immediately. Since these are the Auriga simulations most similar to the Milky Way in their bulge and inner halo chemokinematics, this furthers the evidence for a link between GES and the Milky Way's bar.

In the right panel of Fig. \ref{otherau}, we show histograms for the bar formation times of these 9 GES-like galaxies and for the other 17 bar forming Auriga halos that do not contain a GES-analogue. Not only do the galaxies with a GES-like merger tend to form bars earlier, shown by their median bar formation time being $\sim3\,\mathrm{Gyr}$ earlier than those without such mergers, but the 4 oldest bars in the whole suite of Auriga simulations are formed immediately following a GES-like merger. This suggests that bars older than $8\,\mathrm{Gyr}$ in particular may be very strongly linked to mergers with similar properties to GES. If the Milky Way's bar formed consistent with the older estimates for its age, this would suggest a causal link of GES triggering its bar formation even based on trends alone. It is, however, worth noting that \citet{2020Ro} and \citet{2023Kh} both find in cosmological simulations that strong bars at $z=0$ tend to be correlated with earlier disc formation. Therefore, since early GES-like mergers also require an early disc, the earliest bars in Auriga may be a result of early discs in combination with their GES-like mergers. The Milky Way itself has an early forming disc \citep{2019Ma} in addition to its GES merger event, making it consistent with this scenario.

\subsection{GES and the growth of the Milky Way's disc}
\label{stunt}

\begin{figure}
\centering
\includegraphics[width=0.9\columnwidth]{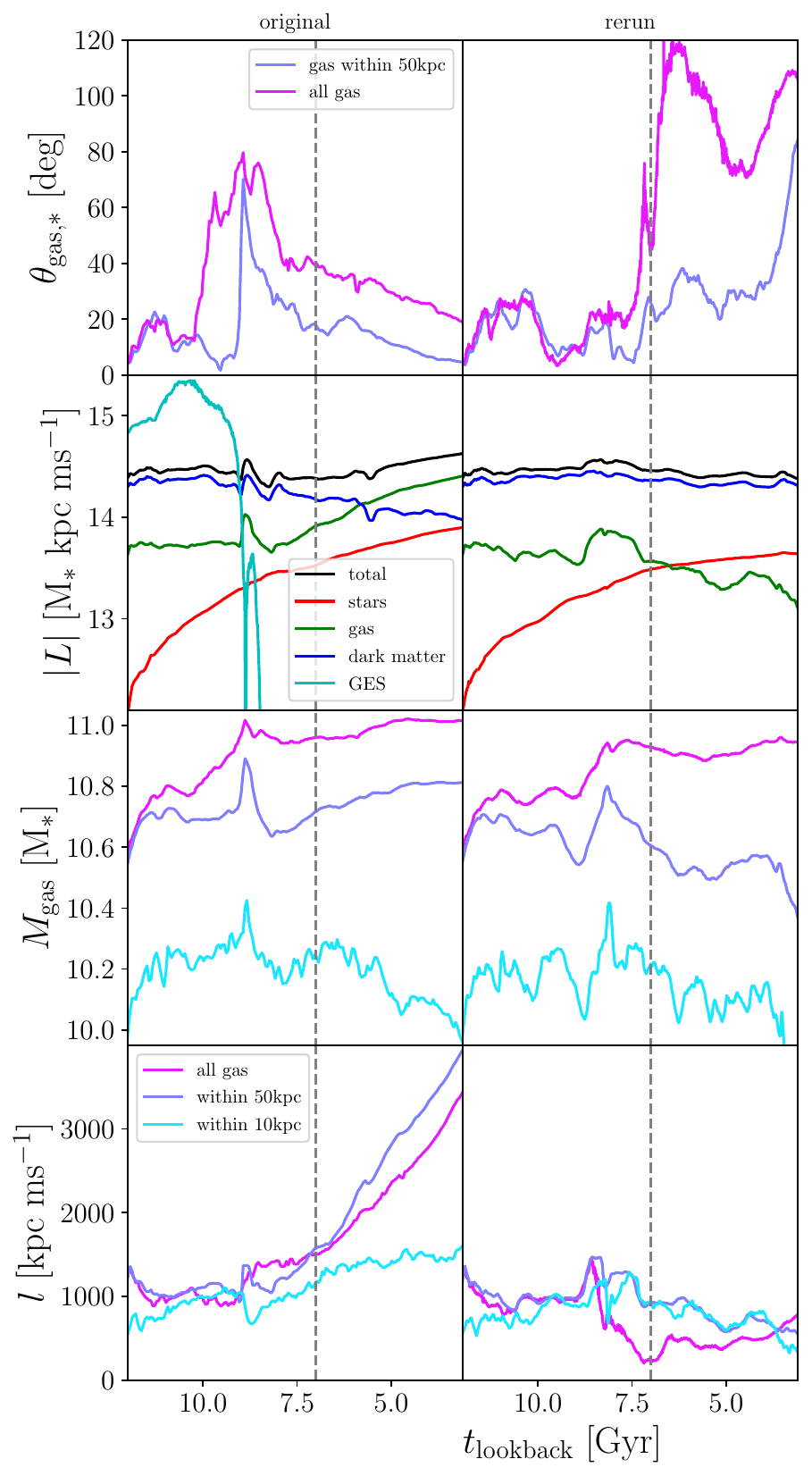}
\caption{Details of the evolution of angular momentum in the original (left panels) and the rerun (right panels) of Au-18 from $12-3\,\mathrm{Gyr}$ lookback time. The grey dashed line shows the time at which the extent of the disc in the rerun diverges from that of the original. Top: The angle in degrees between the stars' and gas' angular momentum vectors within $50\,\mathrm{kpc}$ of the galaxy's centre (blue) and the same angle but taking all gas in the main halo (pink). Second: The magnitude of angular momentum of stars, gas, dark matter, and the total vector, all within 50kpc. For the original, we also plot in cyan the angular momentum of GES-18's bulk motion around the main halo until it merges in the tree. Third: The mass of all gas in the main halo and contained within spheres of $50\,\mathrm{kpc}$ and $10\,\mathrm{kpc}$. Bottom: The specific angular momentum of the gas in the main halo and for the same spheres as above.}
\label{angmom}
\end{figure}

Aside from their bars, Au-18 and our rerun produce very different galaxies by redshift 0 as a result of the presence/absence of GES-18. As such, these shed light on the possible galaxy-wide impact a GES-like merger can have on a galaxy with an otherwise quiet merger history, including our own Milky Way.

As shown in Fig. \ref{newsum}, from $8\,\mathrm{Gyr}$ ago to the present, Au-18's disc grows consistently, while the rerun's disc stagnates and stops growing in extent after a lookback time of $7\,\mathrm{Gyr}$. Even though GES-18 brings in a significant amount of gas to the main halo, the figure's second panel shows that the stellar mass of both discs are almost equal by the end of the simulation. Thus, it is unlikely that a mass difference is enough to account for the difference in disc morphology.

When GES-18 merges with Au-18, it also imparts a large amount of angular momentum to the galaxy. Fig. \ref{angmom} illustrates the evolution until $3\,\mathrm{Gyr}$ ago of angular momentum in the original simulation (left) and the rerun (right). The top panels show the angle separating the stars' spin axis and the gas' spin axis for gas within the inner $50\,\mathrm{kpc}$ and the wider halo. The second panel gives the magnitude of angular momentum associated with each type of matter and with GES-18 and both versions of Au-18 as a whole. The third panel shows the mass of gas within spheres of different radii and the final panel shows the specific angular momentum of the gas within the same spheres.

The angular momentum imparted by GES-18 is of a similar magnitude to the pre-existing angular momentum of the main halo and, as shown in the second left panel of Fig. \ref{angmom}, an order of magnitude greater than the angular momentum in the inner disc-forming regions. Rather than increasing the angular momentum of the main halo, the immediate effect of this is a change in its axis due to GES-18 approaching at an angle of $\sim70^{\circ}$ to the stellar disc. The stellar disc's orientation is shifted by $\sim65^{\circ}$ between lookback times of $9\,\mathrm{Gyr}$ and $8\,\mathrm{Gyr}$. This change in orientation is matched in the gas component too, with the angle between the stellar disc and the gas at all radii returning to near-alignment after the initial perturbation in the top-left panel. After the merger, the angular momentum of the gas increases due to accretion of gas from the outer regions of the halo and beyond. The gas' specific angular momentum sees a dramatic increase outside of $10\,\mathrm{kpc}$ in the lower-left panel, allowing the stellar disc to be built outwards by this gas.

Without GES-18, the inner $50\,\mathrm{kpc}$ of the rerun's main halo, including the stellar disc, do not experience such a change in orientation. There is, however, a change in the angular momentum of the outer regions at $\sim7\,\mathrm{Gyr}$ ago as the surrounding matter forces the vector to almost the same direction as GES-18 did in the original. This results in a stellar disc and inner region of gas which are completely misaligned with the outer gas' rotation as shown in the top-right panel of Fig. \ref{angmom}. In addition to star formation outpacing gas accretion during this time, the newly accreted gas cannot increase the angular momentum of the inner gas due to the misalignment, resulting in a low specific angular momentum (bottom-right panel) and a decreasing total angular momentum (second right panel). As such, the disc does not increase its extent and halts its growth.

This mechanism is responsible for the stagnating disc extent in the rerun for lookback times of $7-4\,\mathrm{Gyr}$. For later times than this, another effect takes over. At $4\,\mathrm{Gyr}$ ago, a starburst triggered by interaction with a smaller galaxy almost fully depletes the gas in the rerun. This allows the still accreting gas from the misaligned outer regions to form a new disc of gas in its own orientation. For the last $2\,\mathrm{Gyr}$ of the simulation, this forms a young polar disc in addition to the main older disc which has ceased star formation. The polar disc's misalignment and the older disc's quenching lead to no further growth of the galaxy's extent, continuing the trend begun at $7\,\mathrm{Gyr}$.

While the development of the polar disc is a relatively indirect effect of removing GES-18, being dependent on a later interaction, the initial impact of the realignment is critical to Au-18's development. If the Milky Way's own GES merged in a similar way, then a similarly large impact may have been made on the Milky Way and be responsible for the entire structure of the galaxy we live in today. We already expect GES to have had a profound impact on the Milky Way's structure. \citet{2018Ha,2019DM,2020Be,2020Gr2} find that the thick disc and non-rotating 'Splash' component are the direct result of an early merger, most likely GES, and are made of both the kinematically heated remnants of the Milky Way's original disc and the stars formed by the gas accreted with GES. Using the Auriga simulations, \citet{2018Gr} and \citet{2020Gr2} then suggest that the thin disc formed from later gas accretion after the end of the merger event. This is consistent with the process described here, with the added possibility that the thin disc's formation may have been out of plane with the early disc if GES did not alter the orientation of what became the thick disc.

\section{Conclusions}

In this work we aimed to explore the relation between the last significant merger of the Milky Way, the Gaia Enceladus/Sausage, and the formation of the Milky Way's bar. To this aim, we used a cosmological zoom-in simulation of a Milky Way-like analogue from the Auriga suite, halo 18 (Au-18), to examine in detail the time of bar formation, and the effects of the concurrent GES-like merger on the host galaxy. To further compare these effects, we ran a controlled cosmological zoom-in simulation using initial conditions of the Au-18 simulation at $z=4$ but with the GES-like halo's progenitors removed. The main halos of the two simulations are near-identical up until the GES-like merger event. In the rerun, there is no GES-like merger, leaving only a more minor merger event (less massive by a factor of 4) at $8.2\,\mathrm{Gyr}$ rather than the original's earlier lookback time of $8.9\,\mathrm{Gyr}$.

Our conclusions are as follows:

\begin{itemize}
  \item We find from the original simulation, that the GES-analogue creates large tidal forces in the host galaxy on its first pericentre. These lead to a perturbation in the bar-like $A_2$ Fourier mode in the outer regions, which propagates inwards to form a bar. Simultaneously, the gas already present in the host galaxy migrates inwards to induce a burst of star formation in the inner regions of the disc, adding mass to the thick disc and contributing to the baryon-dominance of the disc.
  \item In our controlled rerun simulation, these effects are significantly muted without the influence of the main GES-like merger. The later low-mass merger is only able to induce a fluctuating $A_2$ perturbation which fades after $1-2\,\mathrm{Gyr}$. In total, the rerun's bar formation is delayed by over $2\,\mathrm{Gyr}$, and appears to form secularly rather than by the mechanism described for the original.
  \item Taken together, this leads us to conclude that the bar formation in Au-18 is triggered due to the tidal effects of its merger with GES-18.
  \item We also show that GES-18 had a profound impact on the wider disc of Au-18, by bringing its angular momentum into alignment with its surroundings and allowing the accretion of new gas to extend the disc. Our rerun shows the potential effects of an extremely quiescent merger history like the Milky Way's without a significant final merger to connect the galaxy with its surroundings.
  \item We explore the connection between bar formation and GES-like mergers in other Auriga halos which experience such mergers, according to \citet{2019Fa}. We find that the lower mass examples of the mergers which leave a GES-like imprint on the current stellar halo, including Au-18, tend to form bars within $1\,\mathrm{Gyr}$ of their mergers. These also represent the oldest bars in the full simulation suite.
\end{itemize}

Au-18 is a close analogue to the Milky Way in its accretion history and structure, containing both a last significant merger closely resembling GES and an old bar originating around the same time. As such, we have presented a possible path for the formation of the Milky Way's bar being caused by the GES merger, through tidal interactions and the rapid build-up of mass in the inner galaxy. In a forthcoming study, we will search for observational predictions for this scenario from the chemo-kinematics of stars, to uncover bar formation in the Milky Way and other galaxies.

\section*{Acknowledgements}

The authors thank the referee for their helpful and constructive comments. We are very grateful to Prof. Daisuke Kawata for insightful discussions that provided critical motivation for this work. RG is supported by an STFC Ernest Rutherford Fellowship (ST/W003643/1). The simulations presented in this paper were run on the Freya computer cluster at the Max Planck Institute for Astrophysics and the cosma machine operated by the Institute for Computational Cosmology on behalf of the STFC DiRAC HPC Facility ‘www.dirac.ac.uk’.

\section*{Data Availability}

The data presented in this article will be shared on reasonable request to the corresponding author.



\bibliographystyle{mnras}


\begin{thebibliography}{}
\makeatletter
\relax
\def\mn@urlcharsother{\let\do\@makeother \do\$\do\&\do\#\do\^\do\_\do\%\do\~}
\def\mn@doi{\begingroup\mn@urlcharsother \@ifnextchar [ {\mn@doi@}
  {\mn@doi@[]}}
\def\mn@doi@[#1]#2{\def\@tempa{#1}\ifx\@tempa\@empty \href
  {http://dx.doi.org/#2} {doi:#2}\else \href {http://dx.doi.org/#2} {#1}\fi
  \endgroup}
\def\mn@eprint#1#2{\mn@eprint@#1:#2::\@nil}
\def\mn@eprint@arXiv#1{\href {http://arxiv.org/abs/#1} {{\tt arXiv:#1}}}
\def\mn@eprint@dblp#1{\href {http://dblp.uni-trier.de/rec/bibtex/#1.xml}
  {dblp:#1}}
\def\mn@eprint@#1:#2:#3:#4\@nil{\def\@tempa {#1}\def\@tempb {#2}\def\@tempc
  {#3}\ifx \@tempc \@empty \let \@tempc \@tempb \let \@tempb \@tempa \fi \ifx
  \@tempb \@empty \def\@tempb {arXiv}\fi \@ifundefined
  {mn@eprint@\@tempb}{\@tempb:\@tempc}{\expandafter \expandafter \csname
  mn@eprint@\@tempb\endcsname \expandafter{\@tempc}}}

\bibitem[\protect\citeauthoryear{Athanassoula}{Athanassoula}{2002}]{2002At}
Athanassoula E.,  2002, arXiv preprint astro-ph/0209438

\bibitem[\protect\citeauthoryear{Athanassoula}{Athanassoula}{2003}]{2003At}
Athanassoula E.,  2003, \mnras, 341, 1179

\bibitem[\protect\citeauthoryear{Athanassoula \& Misiriotis}{Athanassoula \&
  Misiriotis}{2002}]{2002At2}
Athanassoula E.,  Misiriotis A.,  2002, \mnras, 330, 35

\bibitem[\protect\citeauthoryear{Athanassoula, Rodionov, Peschken  \&
  Lambert}{Athanassoula et~al.}{2016}]{2016At}
Athanassoula E.,  Rodionov S.,  Peschken N.,   Lambert J.,  2016, \apjl, 821,
  90

\bibitem[\protect\citeauthoryear{Baba \& Kawata}{Baba \& Kawata}{2020}]{2020Ba}
Baba J.,  Kawata D.,  2020, Monthly Notices of the Royal Astronomical Society,
  492, 4500

\bibitem[\protect\citeauthoryear{Belokurov, Erkal, Evans, Koposov  \&
  Deason}{Belokurov et~al.}{2018}]{2018Be}
Belokurov V.,  Erkal D.,  Evans N.,  Koposov S.,   Deason A.,  2018, \mnras,
  478, 611

\bibitem[\protect\citeauthoryear{Belokurov, Sanders, Fattahi, Smith, Deason,
  Evans  \& Grand}{Belokurov et~al.}{2020}]{2020Be}
Belokurov V.,  Sanders J.~L.,  Fattahi A.,  Smith M.~C.,  Deason A.~J.,  Evans
  N.~W.,   Grand R.~J.,  2020, Monthly Notices of the Royal Astronomical
  Society, 494, 3880

\bibitem[\protect\citeauthoryear{Bi, Shlosman  \& Romano-D{\'\i}az}{Bi
  et~al.}{2022}]{2022Bi}
Bi D.,  Shlosman I.,   Romano-D{\'\i}az E.,  2022, \apjl, 934, 52

\bibitem[\protect\citeauthoryear{Bissantz \& Gerhard}{Bissantz \&
  Gerhard}{2002}]{2002Bi}
Bissantz N.,  Gerhard O.,  2002, \mnras, 330, 591

\bibitem[\protect\citeauthoryear{Bovy, Leung, Hunt, Mackereth,
  Garc{\'\i}a-Hern{\'a}ndez  \& Roman-Lopes}{Bovy et~al.}{2019}]{2019Bo}
Bovy J.,  Leung H.~W.,  Hunt J.~A.,  Mackereth J.~T.,
  Garc{\'\i}a-Hern{\'a}ndez D.~A.,   Roman-Lopes A.,  2019, \mnras, 490, 4740

\bibitem[\protect\citeauthoryear{Brook, Kawata, Gibson  \& Flynn}{Brook
  et~al.}{2002}]{2002Br}
Brook C.~B.,  Kawata D.,  Gibson B.~K.,   Flynn C.,  2002, \apjl, 585, L125

\bibitem[\protect\citeauthoryear{Brook, Richard, Kawata, Martel  \&
  Gibson}{Brook et~al.}{2007}]{2007Br}
Brook C.,  Richard S.,  Kawata D.,  Martel H.,   Gibson B.~K.,  2007, \apjl,
  658, 60

\bibitem[\protect\citeauthoryear{Chaplin et~al.,}{Chaplin
  et~al.}{2020}]{2020Ch}
Chaplin W.~J.,  et~al., 2020, Nature Astronomy, 4, 382

\bibitem[\protect\citeauthoryear{Cole \& Weinberg}{Cole \&
  Weinberg}{2002}]{2002Co}
Cole A.~A.,  Weinberg M.~D.,  2002, \apjl, 574, L43

\bibitem[\protect\citeauthoryear{Conte et~al.,}{Conte et~al.}{2023}]{2023Co}
Conte Z. A.~L.,  et~al., 2023, arXiv preprint arXiv:2309.10038

\bibitem[\protect\citeauthoryear{Costantin et~al.,}{Costantin
  et~al.}{2023}]{2023Cs}
Costantin L.,  et~al., 2023, Nature, 623, 499

\bibitem[\protect\citeauthoryear{{Deason}, {Belokurov}  \& {Sanders}}{{Deason}
  et~al.}{2019}]{2019De}
{Deason} A.~J.,  {Belokurov} V.,   {Sanders} J.~L.,  2019, \mn@doi [\mnras]
  {10.1093/mnras/stz2793}, \href
  {https://ui.adsabs.harvard.edu/abs/2019MNRAS.490.3426D} {490, 3426}

\bibitem[\protect\citeauthoryear{Di~Matteo, Haywood, Lehnert, Katz, Khoperskov,
  Snaith, G{\'o}mez  \& Robichon}{Di~Matteo et~al.}{2019a}]{2019Di}
Di~Matteo P.,  Haywood M.,  Lehnert M.,  Katz D.,  Khoperskov S.,  Snaith O.,
  G{\'o}mez A.,   Robichon N.,  2019a, Astronomy and Astrophysics-A\&A, 632, A4

\bibitem[\protect\citeauthoryear{Di~Matteo, Haywood, Lehnert, Katz, Khoperskov,
  Snaith, G{\'o}mez  \& Robichon}{Di~Matteo et~al.}{2019b}]{2019DM}
Di~Matteo P.,  Haywood M.,  Lehnert M.,  Katz D.,  Khoperskov S.,  Snaith O.,
  G{\'o}mez A.,   Robichon N.,  2019b, Astronomy \& Astrophysics, 632, A4

\bibitem[\protect\citeauthoryear{Elmegreen, Sundin, Elmegreen  \&
  Sundelius}{Elmegreen et~al.}{1991}]{1991El}
Elmegreen D.,  Sundin M.,  Elmegreen B.,   Sundelius B.,  1991, Astronomy and
  Astrophysics (ISSN 0004-6361), vol. 244, no. 1, April 1991, p. 52-63.
  Research supported by Magn. Bergvalls Stiftelse., 244, 52

\bibitem[\protect\citeauthoryear{Erwin}{Erwin}{2018}]{2018Er}
Erwin P.,  2018, \mnras, 474, 5372

\bibitem[\protect\citeauthoryear{{Fattahi} et~al.,}{{Fattahi}
  et~al.}{2019}]{2019Fa}
{Fattahi} A.,  et~al., 2019, \mn@doi [\mnras] {10.1093/mnras/stz159}, \href
  {https://ui.adsabs.harvard.edu/abs/2019MNRAS.484.4471F} {484, 4471}

\bibitem[\protect\citeauthoryear{{Faucher-Gigu{\`e}re}, {Lidz}, {Zaldarriaga}
  \& {Hernquist}}{{Faucher-Gigu{\`e}re} et~al.}{2009}]{FG09}
{Faucher-Gigu{\`e}re} C.-A.,  {Lidz} A.,  {Zaldarriaga} M.,   {Hernquist} L.,
  2009, \mn@doi [\apj] {10.1088/0004-637X/703/2/1416}, \href
  {http://adsabs.harvard.edu/abs/2009ApJ...703.1416F} {703, 1416}

\bibitem[\protect\citeauthoryear{Feuillet, Feltzing, Sahlholdt  \&
  Casagrande}{Feuillet et~al.}{2020}]{2020Fe}
Feuillet D.~K.,  Feltzing S.,  Sahlholdt C.~L.,   Casagrande L.,  2020, \mnras,
  497, 109

\bibitem[\protect\citeauthoryear{Fragkoudi et~al.,}{Fragkoudi
  et~al.}{2020}]{2020Fr}
Fragkoudi F.,  et~al., 2020, \mnras, 494, 5936

\bibitem[\protect\citeauthoryear{Gallart, Bernard, Brook, Ruiz-Lara, Cassisi,
  Hill  \& Monelli}{Gallart et~al.}{2019}]{2019Ga}
Gallart C.,  Bernard E.~J.,  Brook C.~B.,  Ruiz-Lara T.,  Cassisi S.,  Hill V.,
    Monelli M.,  2019, Nature Astronomy, 3, 932

\bibitem[\protect\citeauthoryear{Gerin, Combes  \& Athanassoula}{Gerin
  et~al.}{1990}]{1990Ge}
Gerin M.,  Combes F.,   Athanassoula E.,  1990, Astronomy and Astrophysics,
  230, 37

\bibitem[\protect\citeauthoryear{Ghosh, Saha, Di~Matteo  \& Combes}{Ghosh
  et~al.}{2021}]{2021Gh}
Ghosh S.,  Saha K.,  Di~Matteo P.,   Combes F.,  2021, \mnras, 502, 3085

\bibitem[\protect\citeauthoryear{Grady, Belokurov  \& Evans}{Grady
  et~al.}{2020}]{2020Gr}
Grady J.,  Belokurov V.,   Evans N.,  2020, \mnras, 492, 3128

\bibitem[\protect\citeauthoryear{Grand et~al.,}{Grand et~al.}{2017}]{2017Gr}
Grand R.~J.,  et~al., 2017, \mnras, 467, 179

\bibitem[\protect\citeauthoryear{Grand et~al.,}{Grand et~al.}{2018}]{2018Gr}
Grand R.~J.,  et~al., 2018, Monthly Notices of the Royal Astronomical Society,
  474, 3629

\bibitem[\protect\citeauthoryear{Grand et~al.,}{Grand et~al.}{2020}]{2020Gr2}
Grand R.~J.,  et~al., 2020, \mnras, 497, 1603

\bibitem[\protect\citeauthoryear{Grand, Fragkoudi, G{\'o}mez, Jenkins,
  Marinacci, Pakmor  \& Springel}{Grand et~al.}{2024}]{2024Gr}
Grand R.~J.,  Fragkoudi F.,  G{\'o}mez F.~A.,  Jenkins A.,  Marinacci F.,
  Pakmor R.,   Springel V.,  2024, arXiv preprint arXiv:2401.08750

\bibitem[\protect\citeauthoryear{Guedes, Mayer, Carollo  \& Madau}{Guedes
  et~al.}{2013}]{2013Gu}
Guedes J.,  Mayer L.,  Carollo M.,   Madau P.,  2013, \apjl, 772, 36

\bibitem[\protect\citeauthoryear{Guo et~al.,}{Guo et~al.}{2023}]{2022Gu}
Guo Y.,  et~al., 2023, The Astrophysical Journal Letters, 945, L10

\bibitem[\protect\citeauthoryear{Haywood, Di~Matteo, Lehnert, Snaith,
  Khoperskov  \& G{\'o}mez}{Haywood et~al.}{2018}]{2018Ha}
Haywood M.,  Di~Matteo P.,  Lehnert M.,  Snaith O.,  Khoperskov S.,   G{\'o}mez
  A.,  2018, The Astrophysical Journal, 863, 113

\bibitem[\protect\citeauthoryear{Helmi, Babusiaux, Koppelman, Massari,
  Veljanoski  \& Brown}{Helmi et~al.}{2018}]{2018He}
Helmi A.,  Babusiaux C.,  Koppelman H.~H.,  Massari D.,  Veljanoski J.,   Brown
  A.~G.,  2018, Nature, 563, 85

\bibitem[\protect\citeauthoryear{Hilmi et~al.,}{Hilmi et~al.}{2020}]{2020Hi}
Hilmi T.,  et~al., 2020, \mnras, 497, 933

\bibitem[\protect\citeauthoryear{Hohl}{Hohl}{1971}]{1971Ho}
Hohl F.,  1971, \apjl, 168, 343

\bibitem[\protect\citeauthoryear{Kalnajs}{Kalnajs}{1972}]{1972Ka}
Kalnajs A.~J.,  1972, Astrophysical Journal, vol. 175, p. 63, 175, 63

\bibitem[\protect\citeauthoryear{Khoperskov, Minchev, Steinmetz, Ratcliffe,
  Walcher  \& Libeskind}{Khoperskov et~al.}{2023}]{2023Kh}
Khoperskov S.,  Minchev I.,  Steinmetz M.,  Ratcliffe B.,  Walcher J.~C.,
  Libeskind N.,  2023, arXiv preprint arXiv:2309.07321

\bibitem[\protect\citeauthoryear{Kraljic, Bournaud  \& Martig}{Kraljic
  et~al.}{2012}]{2012Kr}
Kraljic K.,  Bournaud F.,   Martig M.,  2012, \apjl, 757, 60

\bibitem[\protect\citeauthoryear{Lane, Bovy  \& Mackereth}{Lane
  et~al.}{2023}]{2023La}
Lane J.~M.,  Bovy J.,   Mackereth J.~T.,  2023, \mnras, 526, 1209

\bibitem[\protect\citeauthoryear{{\L}okas}{{\L}okas}{2018}]{2018Lo}
{\L}okas E.~L.,  2018, \apjl, 857, 6

\bibitem[\protect\citeauthoryear{{\L}okas, Athanassoula, Debattista, Valluri,
  Pino, Semczuk, Gajda  \& Kowalczyk}{{\L}okas et~al.}{2014}]{2014Lo}
{\L}okas E.,  Athanassoula E.,  Debattista V.~P.,  Valluri M.,  Pino A.~d.,
  Semczuk M.,  Gajda G.,   Kowalczyk K.,  2014, \mnras, 445, 1339

\bibitem[\protect\citeauthoryear{Lopez-Corredoira, Cabrera-Lavers, Mahoney,
  Hammersley, Garz{\'o}n  \& Gonz{\'a}lez-Fern{\'a}ndez}{Lopez-Corredoira
  et~al.}{2006}]{2006Lo}
Lopez-Corredoira M.,  Cabrera-Lavers A.,  Mahoney T.,  Hammersley P.,
  Garz{\'o}n F.,   Gonz{\'a}lez-Fern{\'a}ndez C.,  2006, The Astronomical
  Journal, 133, 154

\bibitem[\protect\citeauthoryear{Lynden-Bell \& Kalnajs}{Lynden-Bell \&
  Kalnajs}{1972}]{1972Ly}
Lynden-Bell D.,  Kalnajs A.,  1972, \mnras, 157, 1

\bibitem[\protect\citeauthoryear{Mackereth et~al.,}{Mackereth
  et~al.}{2019}]{2019Ma}
Mackereth J.~T.,  et~al., 2019, \mnras, 482, 3426

\bibitem[\protect\citeauthoryear{{Marinacci}, {Pakmor}  \&
  {Springel}}{{Marinacci} et~al.}{2014}]{MPS14}
{Marinacci} F.,  {Pakmor} R.,   {Springel} V.,  2014, \mn@doi [\mnras]
  {10.1093/mnras/stt2003}, \href
  {http://adsabs.harvard.edu/abs/2014MNRAS.437.1750M} {437, 1750}

\bibitem[\protect\citeauthoryear{Martinez-Valpuesta, Aguerri,
  Gonz{\'a}lez-Garc{\'\i}a, Dalla~Vecchia  \& Stringer}{Martinez-Valpuesta
  et~al.}{2017}]{2017Ma}
Martinez-Valpuesta I.,  Aguerri J. A.~L.,  Gonz{\'a}lez-Garc{\'\i}a A.~C.,
  Dalla~Vecchia C.,   Stringer M.,  2017, \mnras, 464, 1502

\bibitem[\protect\citeauthoryear{Masters et~al.,}{Masters
  et~al.}{2011}]{2011Ma}
Masters K.~L.,  et~al., 2011, \mnras, 411, 2026

\bibitem[\protect\citeauthoryear{Melvin et~al.,}{Melvin et~al.}{2014}]{2014Me}
Melvin T.,  et~al., 2014, \mnras, 438, 2882

\bibitem[\protect\citeauthoryear{Men{\'e}ndez-Delmestre, Sheth, Schinnerer,
  Jarrett  \& Scoville}{Men{\'e}ndez-Delmestre et~al.}{2007}]{2007En}
Men{\'e}ndez-Delmestre K.,  Sheth K.,  Schinnerer E.,  Jarrett T.~H.,
  Scoville N.~Z.,  2007, \apjl, 657, 790

\bibitem[\protect\citeauthoryear{Miwa \& Noguchi}{Miwa \&
  Noguchi}{1998}]{1998Mi}
Miwa T.,  Noguchi M.,  1998, \apjl, 499, 149

\bibitem[\protect\citeauthoryear{Naidu et~al.,}{Naidu et~al.}{2021}]{2021Na}
Naidu R.~P.,  et~al., 2021, \apjl, 923, 92

\bibitem[\protect\citeauthoryear{Nepal et~al.,}{Nepal et~al.}{2024}]{2024Ne}
Nepal S.,  et~al., 2024, Astronomy \& Astrophysics, 681, L8

\bibitem[\protect\citeauthoryear{Ness et~al.,}{Ness et~al.}{2013}]{2013Ne}
Ness M.,  et~al., 2013, Monthly Notices of the Royal Astronomical Society, 432,
  2092

\bibitem[\protect\citeauthoryear{Noguchi}{Noguchi}{1987}]{1987No}
Noguchi M.,  1987, \mnras, 228, 635

\bibitem[\protect\citeauthoryear{{Nogueras-Lara} et~al.,}{{Nogueras-Lara}
  et~al.}{2020}]{2020NL}
{Nogueras-Lara} F.,  et~al., 2020, \mn@doi [Nature Astronomy]
  {10.1038/s41550-019-0967-9}, \href
  {https://ui.adsabs.harvard.edu/abs/2020NatAs...4..377N} {4, 377}

\bibitem[\protect\citeauthoryear{{Pakmor}, {Marinacci}  \& {Springel}}{{Pakmor}
  et~al.}{2014}]{PMS14}
{Pakmor} R.,  {Marinacci} F.,   {Springel} V.,  2014, \mn@doi [\apjl]
  {10.1088/2041-8205/783/1/L20}, \href
  {http://adsabs.harvard.edu/abs/2014ApJ...783L..20P} {783, L20}

\bibitem[\protect\citeauthoryear{{Pakmor}, {Springel}, {Bauer}, {Mocz},
  {Munoz}, {Ohlmann}, {Schaal}  \& {Zhu}}{{Pakmor} et~al.}{2016}]{2016Pa}
{Pakmor} R.,  {Springel} V.,  {Bauer} A.,  {Mocz} P.,  {Munoz} D.~J.,
  {Ohlmann} S.~T.,  {Schaal} K.,   {Zhu} C.,  2016, \mn@doi [\mnras]
  {10.1093/mnras/stv2380}, \href
  {https://ui.adsabs.harvard.edu/abs/2016MNRAS.455.1134P} {455, 1134}

\bibitem[\protect\citeauthoryear{{Pakmor} et~al.,}{{Pakmor}
  et~al.}{2017}]{PGG17}
{Pakmor} R.,  et~al., 2017, \mn@doi [\mnras] {10.1093/mnras/stx1074}, \href
  {http://adsabs.harvard.edu/abs/2017MNRAS.469.3185P} {469, 3185}

\bibitem[\protect\citeauthoryear{Peters}{Peters}{1975}]{1975Pe}
Peters W.,  1975, \apjl

\bibitem[\protect\citeauthoryear{{Planck Collaboration XVI}}{{Planck
  Collaboration XVI}}{2014}]{2014Pl}
{Planck Collaboration XVI} 2014, A\&A, 571, A16

\bibitem[\protect\citeauthoryear{Prusti et~al.,}{Prusti et~al.}{2016}]{2016Pr}
Prusti T.,  et~al., 2016, Astronomy \& astrophysics, 595, A1

\bibitem[\protect\citeauthoryear{Rey et~al.,}{Rey et~al.}{2023}]{2023Re}
Rey M.~P.,  et~al., 2023, \mnras, 521, 995

\bibitem[\protect\citeauthoryear{Robin, Marshall, Schultheis  \&
  Reyl{\'e}}{Robin et~al.}{2012}]{2012Ro}
Robin A.~C.,  Marshall D.~J.,  Schultheis M.,   Reyl{\'e} C.,  2012, Astronomy
  \& Astrophysics, 538, A106

\bibitem[\protect\citeauthoryear{Rodriguez-Fernandez \&
  Combes}{Rodriguez-Fernandez \& Combes}{2008}]{2008Ro}
Rodriguez-Fernandez N.,  Combes F.,  2008, Astronomy \& Astrophysics, 489, 115

\bibitem[\protect\citeauthoryear{Rosas-Guevara et~al.,}{Rosas-Guevara
  et~al.}{2020}]{2020Ro}
Rosas-Guevara Y.,  et~al., 2020, \mnras, 491, 2547

\bibitem[\protect\citeauthoryear{Rosas-Guevara et~al.,}{Rosas-Guevara
  et~al.}{2022}]{2022Ro}
Rosas-Guevara Y.,  et~al., 2022, \mnras, 512, 5339

\bibitem[\protect\citeauthoryear{Ruchti et~al.,}{Ruchti et~al.}{2015}]{2015Ru}
Ruchti G.,  et~al., 2015, \mnras, 450, 2874

\bibitem[\protect\citeauthoryear{Sanders, Smith  \& Evans}{Sanders
  et~al.}{2019}]{2019Sa}
Sanders J.~L.,  Smith L.,   Evans N.~W.,  2019, \mnras, 488, 4552

\bibitem[\protect\citeauthoryear{Sanders, Kawata, Matsunaga, Sormani, Smith,
  Minniti  \& Gerhard}{Sanders et~al.}{2023}]{2023Sa}
Sanders J.~L.,  Kawata D.,  Matsunaga N.,  Sormani M.~C.,  Smith L.~C.,
  Minniti D.,   Gerhard O.,  2023, arXiv e-prints, pp arXiv--2311

\bibitem[\protect\citeauthoryear{Schaye et~al.,}{Schaye et~al.}{2015}]{2015Sc}
Schaye J.,  et~al., 2015, \mnras, 446, 521

\bibitem[\protect\citeauthoryear{Sch{\"o}del et~al.,}{Sch{\"o}del
  et~al.}{2023}]{2023Sc}
Sch{\"o}del R.,  et~al., 2023, Astronomy and Astrophysics, 672, L8

\bibitem[\protect\citeauthoryear{Sheth et~al.,}{Sheth et~al.}{2008}]{2008Sh}
Sheth K.,  et~al., 2008, \apjl, 675, 1141

\bibitem[\protect\citeauthoryear{Sormani, Binney  \& Magorrian}{Sormani
  et~al.}{2015}]{2015SM}
Sormani M.~C.,  Binney J.,   Magorrian J.,  2015, \mnras, 454, 1818

\bibitem[\protect\citeauthoryear{Sparke \& Sellwood}{Sparke \&
  Sellwood}{1987}]{1987Sp}
Sparke L.~S.,  Sellwood J.,  1987, \mnras, 225, 653

\bibitem[\protect\citeauthoryear{Springel}{Springel}{2010}]{2010Sp}
Springel V.,  2010, \mnras, 401, 791

\bibitem[\protect\citeauthoryear{{Springel} \& {Hernquist}}{{Springel} \&
  {Hernquist}}{2003}]{SH03}
{Springel} V.,  {Hernquist} L.,  2003, \mn@doi [\mnras]
  {10.1046/j.1365-8711.2003.06206.x}, \href
  {http://adsabs.harvard.edu/abs/2003MNRAS.339..289S} {339, 289}

\bibitem[\protect\citeauthoryear{Springel, Pakmor, Zier  \& Reinecke}{Springel
  et~al.}{2021}]{2021Sp}
Springel V.,  Pakmor R.,  Zier O.,   Reinecke M.,  2021, Monthly Notices of the
  Royal Astronomical Society, 506, 2871

\bibitem[\protect\citeauthoryear{Tremaine \& Weinberg}{Tremaine \&
  Weinberg}{1984}]{1984Tr}
Tremaine S.,  Weinberg M.~D.,  1984, \mnras, 209, 729

\bibitem[\protect\citeauthoryear{{Vogelsberger}, {Genel}, {Sijacki}, {Torrey},
  {Springel}  \& {Hernquist}}{{Vogelsberger} et~al.}{2013}]{VGS13}
{Vogelsberger} M.,  {Genel} S.,  {Sijacki} D.,  {Torrey} P.,  {Springel} V.,
  {Hernquist} L.,  2013, \mn@doi [\mnras] {10.1093/mnras/stt1789}, \href
  {http://adsabs.harvard.edu/abs/2013MNRAS.436.3031V} {436, 3031}

\bibitem[\protect\citeauthoryear{Wegg, Gerhard  \& Portail}{Wegg
  et~al.}{2015}]{2015We}
Wegg C.,  Gerhard O.,   Portail M.,  2015, \mnras, 450, 4050

\bibitem[\protect\citeauthoryear{Weinberg}{Weinberg}{1992}]{1992We}
Weinberg M.~D.,  1992, Astrophysical Journal, Part 1 (ISSN 0004-637X), vol.
  384, Jan. 1, 1992, p. 81-94., 384, 81

\bibitem[\protect\citeauthoryear{Weiner \& Sellwood}{Weiner \&
  Sellwood}{1999}]{1999We}
Weiner B.~J.,  Sellwood J.,  1999, \apjl, 524, 112

\bibitem[\protect\citeauthoryear{Wylie, Clarke  \& Gerhard}{Wylie
  et~al.}{2022}]{2022Wy}
Wylie S.~M.,  Clarke J.~P.,   Gerhard O.~E.,  2022, Astronomy \& Astrophysics,
  659, A80

\makeatother
\end{thebibliography}








\bsp	
\label{lastpage}
\end{document}